\documentclass[11pt,a4paper]{article}

\RequirePackage{ifpdf}
\usepackage{amsmath}
\usepackage{mathtools} 

\usepackage{jheppub}
\usepackage{pstricks}
\usepackage[final]{pdfpages}
\usepackage{ifpdf}
\usepackage{slashed}

\usepackage{hyperref}

\usepackage{color}
\usepackage{graphics}

\usepackage{etoolbox} 
\usepackage{fixmath}

\usepackage{notoccite} 

\usepackage{caption}
\usepackage{subcaption}
\usepackage{amsfonts}
\usepackage{multirow}
\usepackage{epstopdf}

\title{Pseudo-scalar Higgs Boson Production at Threshold N$^3$LO and N$^3$LL QCD}

\author[a]{Taushif Ahmed,}
\author[a]{M.C. Kumar,}
\author[b]{Prakash Mathews,}
\author[a]{Narayan Rana}
\author[a]{and V.~Ravindran}

\affiliation[a]{The Institute of Mathematical Sciences, IV Cross Road, CIT Campus, Chennai 600 113, Tamil Nadu, India}
\affiliation[b]{Saha Institute of Nuclear Physics, 1/AF Bidhan Nagar, Kolkata 700 064, West Bengal, India}

\emailAdd{taushif@imsc.res.in}
\emailAdd{mckumar@imsc.res.in}
\emailAdd{prakash.mathews@saha.ac.in}
\emailAdd{rana@imsc.res.in}
\emailAdd{ravindra@imsc.res.in}

\abstract{
We present the first results on the production of pseudo-scalar through gluon fusion at the LHC 
to N$^3$LO in QCD taking into account only soft gluon effects. 
We have used the effective Lagrangian that describes the coupling of pseudo-scalar with the gluons
in the large top quark mass limit.  We have used recently available quantities namely the three 
loop pseudo-scalar form factor and the third order universal soft function in QCD to achieve this.   
Along with the fixed order results, we also present
the process dependent resummation coefficient for threshold resummation to N$^3$LL in QCD.  
Phenomenological impact of these threshold N$^3$LO
corrections to pseudo-scalar production at the LHC is presented and their role to reduce the
renormalisation scale dependence is demonstrated.
} 
\preprint{}

\begin{document}
\allowdisplaybreaks[4]
\unitlength1cm
\maketitle
\flushbottom


\def\D{{\cal D}}
\def\DD{\overline{\cal D}}
\def\g{\overline{\cal G}}
\def\gm{\gamma}
\def\M{{\cal M}}
\def\ep{\epsilon}
\def\epm1{\frac{1}{\epsilon}}
\def\epm2{\frac{1}{\epsilon^{2}}}
\def\epm3{\frac{1}{\epsilon^{3}}}
\def\epm4{\frac{1}{\epsilon^{4}}}
\def\unM{\hat{\cal M}}
\def\ashat{\hat{a}_{s}}
\def\asmur{a_{s}^{2}(\mu_{R}^{2})}
\def\sigbar{{{\overline {\sigma}}}\left(a_{s}(\mu_{R}^{2}), L\left(\mu_{R}^{2}, m_{H}^{2}\right)\right)}
\def\sigbarn{{{{\overline \sigma}}_{n}\left(a_{s}(\mu_{R}^{2}) L\left(\mu_{R}^{2}, m_{H}^{2}\right)\right)}}
\def\unas{ \left( \frac{\hat{a}_s}{\mu_0^{\epsilon}} S_{\epsilon} \right) }
\def\rnM{{\cal M}}
\def\bt{\beta}
\def\cD{{\cal D}}
\def\cC{{\cal C}}
\def\ca{\text{\tiny C}_\text{\tiny A}}
\def\cf{\text{\tiny C}_\text{\tiny F}}
\def\ct{{\red []}}
\def\sv{\text{SV}}
\def\murOmu{\left( \frac{\mu_{R}^{2}}{\mu^{2}} \right)}
\def\bb{b{\bar{b}}}
\def\bt0{\beta_{0}}
\def\bt1{\beta_{1}}
\def\bt2{\beta_{2}}
\def\bt3{\beta_{3}}
\def\gm0{\gamma_{0}}
\def\gm1{\gamma_{1}}
\def\gm2{\gamma_{2}}
\def\gm3{\gamma_{3}}
\def\nn{\nonumber}
\def\l{\left}
\def\r{\right}
\def\F{{\cal F}}

\newcommand{\dis}[1]{\mathbold{#1}}
\newcommand{\overbar}[1]{mkern-1.5mu\overline{\mkern-1.5mu#1\mkern-1.5mu}\mkern
1.5mu}


\section{Introduction}
\setcounter{equation}{0}
\label{sec:intro}

The spectacular discovery of the Higgs boson~\cite{Aad:2012tfa,
  Chatrchyan:2012xdj} at the Large Hadron Collider (LHC) has put the
Standard Model (SM) of elementary particles in the firm footing.  Most
importantly, the mystery of the electroweak symmetry
breaking~\cite{Higgs:1964ia, Higgs:1964pj, Higgs:1966ev,
  Englert:1964et, Guralnik:1964eu} mechanism can now be solved.  The
consistency of the measured decay rates of the Higgs boson to a pair
of vector bosons namely $W^+W^-, ZZ$ and fermions
$b \overline b,\tau \overline \tau$ with the precise predictions of
the SM for the measured Higgs boson mass of $125$ GeV within the
experimental uncertainty~\cite{CMS:yva, ATLAS:2013sla} makes this
discovery very robust. In addition, there is a strong evidence that
the discovered Higgs boson has spin zero and even
parity~\cite{Aad:2013xqa, Khachatryan:2014kca}.  The ongoing 13 TeV
run at LHC will indeed provide further scope to study the properties
of the Higgs boson in great detail.

While the SM is complete in the sense that all of its predictions have
been tested experimentally, the model suffers from various
deficiencies as it can not explain baryon asymmetry in the Universe,
dark matter, neutrino mass etc.  There are several extensions of the
SM, motivated to address these issues.  The minimal version of
Supersymmetric Standard Model (MSSM)~\cite{Rohini:book} is one of the
most elegant extensions of the SM and it addresses the above mentioned
issues.  The Higgs sector of it contains a pair of Higgs doublets
which after symmetry breaking gives two CP even Higgs bosons h, H and
one CP odd (pseudo-scalar) Higgs boson (A) and two charged Higgs
bosons H$^{\pm}$~\cite{Fayet:1974pd, Fayet:1976et, Fayet:1977yc,
  Dimopoulos:1981zb, Sakai:1981gr, Inoue:1982pi, Inoue:1983pp,
  Inoue:1982ej} .  The predicted upper bound on the mass of the
lightest Higgs boson (h) up to three loop level is
consistent~\cite{Martin:2007pg, Harlander:2008ju, Kant:2010tf} with
the recently observed Higgs boson at the LHC.  The efforts to test the
predictions of MSSM or its variants have already been underway and the
current run at the LHC will shed more light on them.  One of them
could be to look for CP odd Higgs boson in the gluon fusion through
heavy fermions as its coupling is appreciable in the small and
moderate $\tan\beta$, the ratio of vacuum expectation values
$v_i, i=1,2$.  In addition, large gluon flux can boost the cross
section.

Since, the leading order production mechanism of the pseudo-scalar of
mass $m_A$ is through heavy quarks, the cross section is not only
proportional to $\tan\beta$ but also square of the strong coupling
constant.  Like the scalar Higgs boson in SM, the leading order
prediction of the pseudo scalar production at the hadron colliders
suffers from large theoretical uncertainties due to renormalisation
scale $\mu_R$ that enters in the strong coupling constant and the
factorisation scale $\mu_F$ in the gluon distribution functions of the
protons. Predictions based on one loop perturbative Quantum
Chromodynamics (pQCD) corrections~\cite{Kauffman:1993nv,
  Djouadi:1993ji, Spira:1993bb, Spira:1995rr} reduce these
uncertainties (in the conventional range with the central scale
$\mu = m_A/2$ and $m_A=200$ GeV) from about 48\% to 35\% while
increasing the LO cross section substantially, by as large as 67\%.
Effective theory approach in the large top quark mass limit provides
an opportunity to go beyond NLO. Such an approach~\cite{Dawson:1990zj,
  Spira:1995rr}in the case of
scalar Higgs boson
production~\cite{Anastasiou:2002yz,Harlander:2002wh,Ravindran:2003um}
turned out to be the most successful one as the finite mass effects at
NNLO level were found to be within 1\%
\cite{Harlander:2009bw,Harlander:2009mq,Pak:2009dg}.  NNLO predictions
for the production of pseudo-scalar at the hadron colliders are
already available~\cite{Harlander:2002vv,Anastasiou:2002wq,
  Ravindran:2003um}.  The NNLO correction increases the NLO cross
section by about 15\% and reduces the scale uncertainties to about
15\%.  Due to large gluon flux at the threshold, namely when the mass
of A approaches to the partonic centre of mass energy, the cross
section is dominated by the presence of soft gluons.  These
contributions often can spoil the reliability of the predictions based
on fixed order perturbative computations.  Resummation of large
logarithms resulting from soft gluons to all orders in the
perturbation theory provides the solution to this problem.  The
systematic predictions based on the next-to-next-to-leading log (NNLL)
resummed result~\cite{Catani:2003zt, Moch:2005ky, Laenen:2005uz, Ravindran:2005vv,
  Ravindran:2006cg,Idilbi:2005ni,Ahrens:2008nc,deFlorian:2009hc,
  Schmidt:2015cea} demonstrate the reliability of the approach and
also reduce the scale uncertainties.

A complete calculation at NNLO~\cite{Anastasiou:2002yz,
  Harlander:2002wh, Ravindran:2003um} and leading logarithms at
N$^{3}$LO in the threshold limit~\cite{Moch:2005ky, Laenen:2005uz, Ravindran:2005vv,
  Ravindran:2006cg, Idilbi:2005ni} and NNLL soft gluon
resummation~\cite{Catani:2003zt} for the scalar Higgs boson production
are known for more than a decade.  Recently there have been series of
works on predicting inclusive scalar Higgs boson production beyond
this level.  The computation of $\delta(1-z)$ contribution at
N$^{3}$LO in the threshold limit~\cite{Anastasiou:2014vaa} was the
first among them.  This was confirmed independently
in~\cite{Li:2014bfa}.  Later on the sub-leading collinear logarithms
were computed in~\cite{Anastasiou:2014lda,deFlorian:2014vta}.  Spin
off of the result presented in~\cite{Anastasiou:2014vaa} is the
computation of N$^{3}$LO prediction for the Drell-Yan
production~\cite{Ahmed:2014cla, Li:2014bfa, Catani:2014uta} at the
hadron colliders in the threshold limit. In addition, one can obtain
N$^{3}$LO threshold corrections to the Higgs boson production through
bottom quark annihilation~\cite{Ahmed:2014cha} and also in association
with vector boson~\cite{Kumar:2014uwa} at the hadron colliders. Later,
along the same direction, rapidity distribution of the Higgs boson in
gluon fusion~\cite{Ahmed:2014uya}, DY~\cite{Ahmed:2014uya} and Higgs
boson in bottom quark annihilation~\cite{Ahmed:2014era} were obtained
at threshold N$^{3}$LO QCD.

A milestone in this direction was achieved by Anastasiou
{\textit{et. al.}} who have now accomplished the complete N$^{3}$LO
prediction~\cite{Anastasiou:2015ema} of the scalar Higgs boson
production through gluon fusion at the hadron colliders in the
effective theory.  These third order corrections increase the cross
section by a few percent, about 2\% and reduce the scale uncertainty
by about 2\%.  Using these predictions, it is now possible to obtain
the soft gluon resummation at N$^{3}$LL, see~\cite{Catani:2014uta,
  Bonvini:2014joa}.

While the next step in the wish list is to obtain the N$^{3}$LO
predictions for the pseudo-scalar production through gluon fusion, the
first task in this direction is to obtain the threshold enhanced cross
section at N$^{3}$LO level.  One of the crucial ingredients is the
form factor of the effective composite operators that couple to
pseudo-scalar, computed between partonic states.  One and two loop
results for them between gluon states were computed for NNLO
production cross
section~\cite{Harlander:2002vv,Anastasiou:2002wq,Ravindran:2004mb},
the analytical results up to two loop level can be found
in~\cite{Ravindran:2004mb}.  These were computed in dimensional
regularisation where the space time dimension is $d=4+\epsilon$.
Threshold corrections to pseudo-scalar production at N$^3$LO level
requires the knowledge of the form factors up to three loop level.  We
also need to know one and two loop corrections computed to desired
accuracy in $\epsilon$, namely up to $\epsilon^2$ for one loop and up
to $\epsilon$ at two loops. In~\cite{Ahmed:2015qpa}, we obtained the
three loop form factors of the effective composite operators between
quark and gluon states at three loop level along with the lower order
ones to desired accuracies in $\epsilon$.  In the present article we
will describe how threshold corrections at N$^3$LO level can be
obtained from the formalism developed in~\cite{Ravindran:2005vv,
  Ravindran:2006cg} using the available information on recently
computed three loop form factor of the pseudo scalar Higgs
boson~\cite{Ahmed:2015qpa}, the universal soft-collinear
distribution~\cite{Ahmed:2014cla} and operator renormalisation
constant~\cite{Larin:1993tq, Zoller:2013ixa, Ahmed:2015qpa} and the
mass factorisation kernels~\cite{Vogt:2004mw, Moch:2004pa} known to
three loop level.  In addition, we compute third order correction to
the $N$-independent part of the resummed cross
section~\cite{Sterman:1986aj,Catani:1989ne} using our
formalism~\cite{Ravindran:2005vv, Ravindran:2006cg}.  We also present
the numerical impact of our findings with a brief conclusion.

The underlying effective theory is discussed in the
Sec.~\ref{sec:Lagrang}. This is followed by a short description of the
formalism which has been employed to compute the soft-plus-virtual
cross section in Sec.~\ref{sec:ThreResu}. We present the analytical
results of these findings in the Sec.~\ref{sec:Res} up to N$^3$LO in
QCD. In Sec.~\ref{sec:Resum}, the N-independent parts of the threshold
resummed cross section in Mellin space have been presented up to third
order in QCD. Before making concluding remarks, in Sec.~\ref{sec:Disc}
we demonstrate the numerical implications of the fixed order
soft-plus-virtual cross sections to N$^3$LO at LHC.


\section{The Effective Lagrangian}
\label{sec:Lagrang}

A pseudo-scalar couples to gluons only indirectly through a virtual
heavy quark loop which can be integrated out in the infinite quark
mass limit.  The effective Lagrangian~\cite{Chetyrkin:1998mw}
describing the interaction between pseudo-scalar $\chi^{A}$ and the
QCD particles in the infinitely large top quark mass limit is given by
\begin{align} {\cal L}^{A}_{\rm eff} = \chi^{A}(x) \Big[ - \frac{1}{8}
  {C}_{G} O_{G}(x) - \frac{1}{2} {C}_{J} O_{J}(x)\Big]
\end{align}
where the two operators are defined as \\
\begin{equation}
  O_{G}(x) = G^{\mu\nu}_a \tilde{G}^{\rho
    \sigma}_a \equiv  \epsilon_{\mu \nu \rho \sigma} G^{\mu\nu}_a G^{\rho
    \sigma}_a\, ,\qquad
  O_{J}(x) = \partial_{\mu} \left( \bar{\psi}
    \gamma^{\mu}\gamma_5 \psi \right)  \,.
  \label{eq:operators}
\end{equation}
The Wilson coefficients $C_{G}$ and $C_{J}$ of the two operators are
the consequences of integrating out the heavy quark loop in effective
theory. $C_{G}$ does not receive any QCD corrections beyond one loop
because of Adler-Bardeen theorem, whereas $C_{J}$ starts only at
second order in the strong coupling constant.  These Wilson
coefficients are given by~\cite{Chetyrkin:1998mw}
\begin{align}
  \label{eq:const}
  C_{G} &= -a_{s} 2^{\frac{5}{4}} G_{F}^{\frac{1}{2}}
          {\rm cot}\beta\,, 
          \nonumber\\
  C_{J} &= - \left[ a_{s} C_{F} \left( \frac{3}{2} - 3\ln
          \frac{\mu_{R}^{2}}{m_{t}^{2}} \right) + a_s^2 C_J^{(2)} + \cdots \right] C_{G}\, .
\end{align}
The symbols $G^{\mu\nu}_{a}$ and $\psi$ represent gluonic field
strength tensor and quark field, respectively. $G_{F}$ stands for the
Fermi constant and ${\rm cot}\beta$ is the mixing angle in the
Two-Higgs-Doublet model. $m_{A}$ and $m_{t}$ symbolise the masses of
the pseudo-scalar and top quark (heavy quark), respectively.  The
strong coupling constant
$a_{s} \equiv a_{s} \left( \mu_{R}^{2} \right)$ is renormalised at the
mass scale $\mu_{R}$ and is related to the unrenormalised one,
${\hat a}_{s} \equiv {\hat g}_{s}^{2}/16\pi^{2}$, through
\begin{align}
  \label{eq:asAasc}
  {\hat a}_{s} S_{\epsilon} = \left( \frac{\mu^{2}}{\mu_{R}^{2}}  \right)^{\epsilon/2}
  Z_{a_{s}} a_{s}
\end{align}
with
$S_{\epsilon} = {\rm exp} \left[ (\gamma_{E} - \ln 4\pi)\epsilon/2
\right]$
and the scale $\mu$ is introduced to keep the unrenormalized strong
coupling constant dimensionless in $d=4+\epsilon$ space-time
dimensions. The renormalisation constant $Z_{a_{s}}$ up to
${\cal O}(a_{s}^{3})$ is given by
\begin{align}
  \label{eq:Zas}
  Z_{a_{s}}&= 1+ a_s\left[\frac{2}{\epsilon} \beta_0\right]
             + a_s^2 \left[\frac{4}{\epsilon^2 } \beta_0^2
             + \frac{1}{\epsilon}  \beta_1 \right]
             + a_s^3 \left[\frac{8}{ \epsilon^3} \beta_0^3
             +\frac{14}{3 \epsilon^2}  \beta_0 \beta_1 +  \frac{2}{3
             \epsilon}   \beta_2 \right]\,.
\end{align}
The coefficient of the QCD $\beta$ function $\beta_{i}$ are given
by~\cite{Tarasov:1980au}
\begin{align}
  \beta_0&={11 \over 3 } C_A - {2 \over 3 } n_f \, ,
           \nonumber \\[0.5ex]
  \beta_1&={34 \over 3 } C_A^2- 2 n_f C_F -{10 \over 3} n_f C_A \, ,
           \nonumber \\[0.5ex]
  \beta_2&={2857 \over 54} C_A^3 
           -{1415 \over 54} C_A^2 n_f
           +{79 \over 54} C_A n_f^2
           +{11 \over 9} C_F n_f^2
           -{205 \over 18} C_F C_A n_f
           + C_F^2 n_f 
\end{align}
with the SU(N) QCD color factors
\begin{equation}
  C_A=N,\quad \quad \quad C_F={N^2-1 \over 2 N}\,.
\end{equation}
$n_f$ is the number of active light quark flavors.

\section{Threshold Corrections}
\label{sec:ThreResu}

The inclusive cross-section for the production of a colorless pseudo
scalar at the hadron colliders can be computed using
\begin{align}
  \sigma^A(\tau,m_A^2) = \sigma^{A,(0)}(\mu_{R}^{2}) \sum_{a,b=q,\bar{q},g} 
  \int_{\tau}^{1} dy ~ \Phi_{ab}(y,\mu_F^2)
  \Delta_{ab}^{A}\left(\frac{\tau}{y}, m_A^2, \mu_R^2, \mu_F^2\right) \,
  \label{eqn.tot1}
\end{align}
where, the born cross section at the parton level including the finite
top mass dependence is given by
\begin{align}
  \sigma^{A,(0)}(\mu_{R}^{2}) = \frac{\pi \sqrt{2} G_F}{16} a_s^2 {\rm
  cot}^2\beta ~\big|  \tau_A f(\tau_A)\big|^2.
\end{align}
Here $\tau_A = 4m_t^2/m_A^2$ and the function $f(\tau_{A})$ is given
by
\begin{align}
  \label{eq:FA}
  &f(\tau_{A}) = 
    \begin{cases}
      {\rm arcsin}^{2}\frac{1}{\sqrt{\tau_{A}}} & \tau_{A} \geq 1\,, \\
      -\frac{1}{4} \left( \ln
        \frac{1-\sqrt{1-\tau_{A}}}{1+\sqrt{1-\tau_{A}}} +i \pi
      \right)^{2} & \tau_{A} < 1\,.
    \end{cases}
\end{align} 
while the parton flux is given by
\begin{eqnarray}
  \Phi_{ab}(y,\mu_F^2) = \int_y^1 {dx \over x} f_a(x,\mu_F^2) f_b\left({y \over x},\mu_F^2\right) \, ,
\end{eqnarray}
where, $f_a$ and $f_b$ are the parton distribution functions (PDFs) of
the initial state partons $a$ and $b$, renormalised at the
factorisation scale $\mu_{F}$. Here,
${\Delta}^{A}_{a b} \left(\frac{\tau}{y}, m_A^2, \mu_{R}^{2},
  \mu_F^2\right)$
are the partonic level cross sections, for the subprocess initiated by
the partons $a$ and $b$, computed after performing the overall
operator UV renormalisation at scale $\mu_{R}$ and mass factorisation
at a scale $\mu_{F}$. The variable $\tau$ is defined as $q^{2}/s$ with
$q^{2} = m_{A}^{2}$.

The goal of this article is to study the impact of the soft gluon
contributions to the pseudo-scalar production cross section at hadron
colliders. The infrared safe contribution is obtained by adding the
soft part of the cross section to the ultraviolet (UV) renormalised
virtual part and performing the mass factorisation using appropriate
counter terms. This combination is often called the soft-plus-virtual
(SV) cross section whereas the remaining portion is known as hard
part. Thus, we write the partonic cross section as
\begin{align}
  \label{eq:PartsOfDelta}
  &{\Delta}^{A}_{a b} (z, q^{2}, \mu_{R}^{2}, \mu_F^2) 
    = {\Delta}^{A, \text{SV}}_{a b} (z, q^{2}, \mu_{R}^{2}, \mu_F^2)  
    + {\Delta}^{A,\text{hard}}_{a b} (z, q^{2}, \mu_{R}^{2}, \mu_F^2) \, 
\end{align}
with
$ z \equiv q^{2}/\hat{s} = \tau/(x_{1} x_{2}) \,$.  The threshold
contributions
${\Delta}^{A,\text{SV}}_{a b} (z, q^{2}, \mu_{R}^{2}, \mu_F^2)$
contains only the distributions of kind $\delta(1-z)$ and
${\cal{D}}_{i}$, where the latter one is defined through
\begin{align}
  \label{eq:calD}
  {\cal{D}}_{i} \equiv \left[ \frac{\ln^{i}(1-z)}{1-z} \right]_{+}\, .
\end{align}
On the other hand, the hard part ${\Delta}^{A,\text{hard}}_{a b}$
contains all the terms regular in $z$. The SV cross-section in
$z$-space is computed in $d=4+\ep$ dimensions, as formulated for the
first time in \cite{Ravindran:2005vv, Ravindran:2006cg}, using
\begin{align}
  \label{eq:sigma}
  \Delta^{A,\sv}_{g} (z, q^2, \mu_{R}^{2}, \mu_F^2) = 
  {\cal C} \exp \Big( \Psi^A_{g} \left(z, q^2, \mu_R^2, \mu_F^2,
  \epsilon \right)  \Big)  \Big|_{\epsilon = 0}
\end{align}
where, $\Psi^A_{g} \left(z, q^2, \mu_R^2, \mu_F^2, \epsilon \right)$
is a finite distribution and ${\cal C}$ is the Mellin convolution
defined as
\begin{equation}
  \label{eq:conv}
  {\cal C} e^{f(z)} = \delta(1-z) + \frac{1}{1!} f(z) + \frac{1}{2!} f(z) \otimes f(z) + \cdots \, .
\end{equation}
Here $\otimes$ represents Mellin convolution and $f(z)$ is a
distribution of the kind $\delta(1-z)$ and ${\cal D}_i$. The subscript
$g$ signifies the gluon initiated production of the pseudo-scalar. The
equivalent formalism of the SV approximation is in the Mellin (or
$N$-moment) space, where instead of distributions in $z$ the dominant
contributions come from the continuous functions of the variable $N$
(see~\cite{Sterman:1986aj, Catani:1989ne}) and the threshold limit of
$z \rightarrow 1$ is translated to $N \rightarrow \infty$.
The $\Psi^A_{g} \left(z, q^2, \mu_R^2, \mu_F^2, \epsilon \right)$ is
constructed from the form factors
${\F}^A_{g} (\hat{a}_s, Q^2, \mu^2, \epsilon)$ with $Q^{2}=-q^{2}$,
the overall operator UV renormalisation constant
$Z^A_{g}(\hat{a}_s, \mu_R^2, \mu^2, \epsilon)$, the soft-collinear
distribution $\Phi^A_{g}(\hat{a}_s, q^2, \mu^2, z, \epsilon)$ arising
from the real radiations in the partonic subprocesses and the mass
factorisation kernels
$\Gamma_{gg} (\hat{a}_s, \mu_F^2, \mu^2, z, \epsilon)$.  In terms of
the above-mentioned quantities it takes the following form, as
presented in \cite{Ravindran:2006cg, Ahmed:2014cla, Ahmed:2014cha}
\begin{align}
  \label{eq:psi}
  \Psi^{A}_{g} \left(z, q^2, \mu_R^2, \mu_F^2, \epsilon \right)  
  = &\left( \ln \Big[ Z^A_{g} (\hat{a}_s, \mu_R^2, \mu^2, \epsilon) \Big]^2 
      + \ln \Big|  {\F}^A_{g} (\hat{a}_s, Q^2, \mu^2, \epsilon)   \Big|^2 \right) \delta(1-z) 
      \nonumber\\
    & + 2 \Phi^A_{g} (\hat{a}_s, q^2, \mu^2, z, \epsilon) 
      - 2 {\cal C} \ln \Gamma_{gg} (\hat{a}_s, \mu_F^2, \mu^2, z, \epsilon) \, .
\end{align}
In the subsequent sections, we will demonstrate the methodology to get
these ingredients to compute the SV cross section of pseudo-scalar
production at N$^{3}$LO.

\subsection{The Form Factor}
\label{ss:FF}

The quark and gluon form factors represent the QCD loop corrections to
the transition matrix element from an on-shell quark-antiquark pair or
two gluons to a color-neutral operator $O$.  For the pseudo-scalar
production through gluon fusion, we need to consider two operators
$O_{G}$ and $O_{J}$, defined in Eq.~(\ref{eq:operators}), which yield
in total two form factors.
The unrenormalised gluon form factors at ${\cal O}({\hat a}_{s}^{n})$
are defined~\cite{Ahmed:2015qpa} through

\begin{align}
  \label{eq:DefFg}
  {\hat{\cal F}}^{G,(n)}_{g} \equiv \frac{\langle{\hat{\cal
  M}}^{G,(0)}_{g}|{\hat{\cal M}}^{G,(n)}_{g}\rangle}{\langle{\hat{\cal
  M}}^{G,(0)}_{g}|{\hat{\cal M}}^{G,(0)}_{g}\rangle}\, ,
  \qquad \qquad \qquad
  {\hat{\cal F}}^{J,(n)}_{g} \equiv \frac{\langle{\hat{\cal
  M}}^{G,(0)}_{g}|{\hat{\cal M}}^{J,(n+1)}_{g}\rangle}{\langle{\hat{\cal
  M}}^{G,(0)}_{g}|{\hat{\cal M}}^{J,(1)}_{g}\rangle}
\end{align}
where, $n=0, 1, 2, 3, \ldots$\,\,. In the above expressions
$|{\hat{\cal M}}^{\lambda,(n)}_{g}\rangle$ $(\lambda = G,J)$ is the
${\cal O}({\hat a}_{s}^{n})$ contribution to the unrenormalised matrix
element described by the bare operator $[O_{\lambda}]_B$.
In terms of these quantities, the full matrix element and the full
form factors can be written as a series expansion in ${\hat a}_{s}$ as

\begin{align}
  \label{eq:DefFlambda}
  |{\cal M}^{\lambda}_{g}\rangle \equiv \sum_{n=0}^{\infty} {\hat
  a}^{n}_{s} S^{n}_{\epsilon}
  |{\hat{\cal M}}^{\lambda,(n)}_{g} \rangle \, , 
  \qquad \qquad 
  {\cal F}^{\lambda}_{g} \equiv
  \sum_{n=0}^{\infty} \left[ {\hat a}_{s}^{n}
  \left( \frac{Q^{2}}{\mu^{2}} \right)^{n\frac{\epsilon}{2}}
  S_{\epsilon}^{n}  {\hat{\cal F}}^{\lambda,(n)}_{g}\right]\, ,
\end{align}
\\
where $Q^{2}=-2\, p_{1}.p_{2}=-q^{2}$ and $p_i$ ($p_{i}^{2}=0$) are
the momenta of the external on-shell gluons. Note that
$|{\hat{\cal M}}^{J,(n)}_{g}\rangle$ starts at $n=1$ i.e. from one
loop level.

The form factor for the production of a pseudo-scalar through gluon
fusion, ${\hat \F}^{A,(n)}_{g}$, can be written in terms of the two
individual form factors, Eq.~(\ref{eq:DefFlambda}), as follows:
\begin{align}
  \label{eq:FFDefMatEle}
  {\F}^{A}_{g} = {\F}^{G}_{g} + \Bigg(
  \frac{Z_{GJ}}{Z_{GG}} + \frac{4 C_{J}}{C_{G}} \frac{Z_{JJ}}{Z_{GG}} \Bigg)
  {\F}^{J}_{g} \frac{\langle{\hat{\cal
  M}}^{G,(0)}_{g}|{\hat{\cal M}}^{J,(1)}_{g}\rangle}{\langle{\hat{\cal
  M}}^{G,(0)}_{g}|{\hat{\cal M}}^{G,(0)}_{g}\rangle}\,.  
\end{align}
In the above expression, the quantities $Z_{ij} (i,j= G, J)$ are the
overall operator renormalisation constants which are required to
introduce in the context of UV renormalisation.  These are are
discussed in our recent article~\cite{Ahmed:2015qpa} in great
detail. The ingredients of the form factor ${\F}^{A}_{g}$, namely,
${\F}^{G}_{g}$ and ${\F}^{J}_{g}$ have been calculated up to three
loop level by some of us and presented in the same
article~\cite{Ahmed:2015qpa}. Using those results we obtain the three
loop form factor for the pseudo-scalar production through gluon
fusion.
In this section, we present the unrenormalized form factors
${\hat \F}^{A,(n)}_{g}$ up to three loop where the components are
defined through the expansion
\begin{align}
  \label{eq:FF3}
  {\F}^{A}_{g} \equiv
  \sum_{n=0}^{\infty} \left[ {\hat a}_{s}^{n}
  \left( \frac{Q^{2}}{\mu^{2}} \right)^{n\frac{\epsilon}{2}}
  S_{\epsilon}^{n}  {\hat{\F}}^{A,(n)}_{g}\right] \, .
\end{align}
We present the unrenormalized results for the choice of the scale
$\mu_{R}^{2}=\mu_{F}^{2}=q^{2}$\, as follows:
\begin{align}
  \label{eq:FF}
  {\hat \F}^{A,(1)}_{g} &= {\dis{C_{A}}} \Bigg\{ - \frac{8}{\epsilon^2}
                          + 4 +
                          \zeta_2 
                          + 
                          \epsilon \Bigg( - 6 - \frac{7}{3} \zeta_3
                          \Bigg)
                          + \epsilon^2 \Bigg( 7  -
                          \frac{\zeta_2}{2} + \frac{47}{80}  \zeta_2^2
                          \Bigg) 
                          + 
                          \epsilon^3 \Bigg( - \frac{15}{2} 
                          \nonumber\\
                        &+
                          \frac{3}{4} \zeta_2 +  \frac{7}{6} \zeta_3 +
                          \frac{7}{24} \zeta_2 \zeta_3  - 
                          \frac{31}{20} \zeta_5 \Bigg)
                          + \epsilon^4 \Bigg(  \frac{31}{4} -
                          \frac{7}{8} \zeta_2 -  \frac{47}{160}
                          \zeta_2^2 +  
                          \frac{949}{4480} \zeta_2^3 - \frac{7}{4}
                          \zeta_3 -  \frac{49}{144} \zeta_3^2 \Bigg) 
                          \nonumber\\
                        &+  
                          \epsilon^5 \Bigg( - \frac{63}{8} +
                          \frac{15}{16} \zeta_2 +  \frac{141}{320}
                          \zeta_2^2 + \frac{49}{24} \zeta_3 -  
                          \frac{7}{48} \zeta_2 \zeta_3 +
                          \frac{329}{1920} \zeta_2^2 \zeta_3 +
                          \frac{31}{40} \zeta_5 +  
                          \frac{31}{160} \zeta_2 \zeta_5 
                          \nonumber\\
                        &-
                          \frac{127}{112} \zeta_7 \Bigg)
                          + \epsilon^6
                          \Bigg( \frac{127}{16}  - \frac{31}{32}
                          \zeta_2 - \frac{329}{640} \zeta_2^2  - 
                          \frac{949}{8960} \zeta_2^3 +
                          \frac{55779}{716800} \zeta_2^4  -
                          \frac{35}{16} \zeta_3 +  
                          \frac{7}{32} \zeta_2 \zeta_3 
                          \nonumber\\
                        &+ \frac{49}{288}
                          \zeta_3^2 +  \frac{49}{1152} \zeta_2 \zeta_3^2 - 
                          \frac{93}{80} \zeta_5 - \frac{217}{480}
                          \zeta_3 \zeta_5 \Bigg) 
                          +  
                          \epsilon^7 \Bigg( - \frac{255}{32} +
                          \frac{63}{64} \zeta_2 +  \frac{141}{256}
                          \zeta_2^2  + \frac{2847}{17920} \zeta_2^3 
                          \nonumber\\
                        &+ 
                          \frac{217}{96} \zeta_3 - \frac{49}{192}
                          \zeta_2 \zeta_3  - \frac{329}{3840} \zeta_2^2
                          \zeta_3 +  
                          \frac{949}{15360} \zeta_2^3 \zeta_3 -
                          \frac{49}{192} \zeta_3^2 -  \frac{343}{10368}
                          \zeta_3^3 +  
                          \frac{217}{160} \zeta_5 
                          \nonumber\\
                        &- \frac{31}{320}
                          \zeta_2 \zeta_5 +  \frac{1457}{12800}
                          \zeta_2^2 \zeta_5 +  
                          \frac{127}{224} \zeta_7 + \frac{127}{896}
                          \zeta_2 \zeta_7  - \frac{511}{576} \zeta_{9} \Bigg) \Bigg\}\,,
                          \nonumber\\
  {\hat \F}^{A,(2)}_{g} &= {\dis{C_{F} n_{f}}} \Bigg\{ - \frac{80}{3} +
                          6  \ln
                          \left(\frac{q^2}{m_t^2}\right)  + 8
                          \zeta_3 
                          +  \epsilon \Bigg(
                          \frac{2827}{36} - 9  \ln
                          \left(\frac{q^2}{m_t^2}\right)   -
                          \frac{19}{6} \zeta_2 -  
                          \frac{8}{3} \zeta_2^2 
                          \nonumber\\
                        &- \frac{64}{3} \zeta_3
                          \Bigg)  
                          +
                          \epsilon^2 \Bigg(  - \frac{70577}{432} +
                          \frac{21}{2}  \ln
                          \left(\frac{q^2}{m_t^2}\right)   + 
                          \frac{1037}{72} \zeta_2 - \frac{3}{4}
                          \ln \left(\frac{q^2}{m_t^2} \right)
                          \zeta_2  + \frac{64}{9} \zeta_2^2 +
                          \frac{455}{9} \zeta_3  
                          \nonumber\\
                        &- 
                          \frac{10}{3} \zeta_2 \zeta_3 + 8 \zeta_5
                          \Bigg) 
                          + 
                          \epsilon^3 \Bigg( \frac{1523629}{5184} -
                          \frac{45}{4}  \ln
                          \left(\frac{q^2}{m_t^2}\right)  -
                          \frac{14975}{432} \zeta_2  + 
                          \frac{9}{8}  \ln
                          \left(\frac{q^2}{m_t^2}\right) \zeta_2 
                          \nonumber\\
                        &- 
                          \frac{70997}{4320} \zeta_2^2 +  \frac{22}{35}
                          \zeta_2^3  - 
                          \frac{3292}{27} \zeta_3 + \frac{7}{4}
                          \ln \left(\frac{q^2}{m_t^2}\right)
                          \zeta_3  +  \frac{80}{9} \zeta_2 \zeta_3 + 
                          15 \zeta_3^2 - \frac{64}{3} \zeta_5 \Bigg)    
                          \nonumber\\
                        &+ \epsilon^4 \Bigg( -
                          \frac{30487661}{62208}  + \frac{93}{8}   \ln
                          \left( \frac{q^2}{m_t^2}\right) + 
                          \frac{43217}{648} \zeta_2 - \frac{21}{16}
                          \ln \left(\frac{q^2}{m_t^2}\right)
                          \zeta_2   + \frac{1991659}{51840} \zeta_2^2 
                          \nonumber\\
                        &- 
                          \frac{141}{320}  \ln \left(\frac{q^2}{m_t^2}\right) \zeta_2^2 -
                          \frac{176}{105} \zeta_2^3  +
                          \frac{694231}{2592} \zeta_3  - 
                          \frac{21}{8}  \ln \left(\frac{q^2}{m_t^2}\right) \zeta_3 -
                          \frac{9757}{432} \zeta_2 \zeta_3  - 
                          \frac{1681}{180} \zeta_2^2 \zeta_3 
                          \nonumber\\
                        &- 40
                          \zeta_3^2 + \frac{8851}{180} \zeta_5  - 
                          2 \zeta_2 \zeta_5 - \frac{127}{8} \zeta_7
                          \Bigg) \Bigg\} 
                          + 
                          {\dis{C_{A} n_{f}}} \Bigg\{  -
                          \frac{8}{3 \epsilon^3}  + \frac{20}{9
                          \epsilon^2} + 
                          \Bigg( \frac{106}{27} + 2 \zeta_2 \Bigg)
                          \frac{1}{\epsilon} 
                          \nonumber\\
                        &- \frac{1591}{81} -
                          \frac{5}{3} \zeta_2     - \frac{74}{9}
                          \zeta_3 
                          + 
                          \epsilon \Bigg( \frac{24107}{486} -
                          \frac{23}{18} \zeta_2  + \frac{51}{20}
                          \zeta_2^2  + \frac{383}{27} \zeta_3 \Bigg) 
                          + 
                          \epsilon^2 \Bigg( - \frac{146147}{1458} 
                          \nonumber\\
                        &+
                          \frac{799}{108} \zeta_2  - \frac{329}{72} \zeta_2^2 - 
                          \frac{1436}{81} \zeta_3 + \frac{25}{6}
                          \zeta_2 \zeta_3  - \frac{271}{30} \zeta_5
                          \Bigg) 
                          + 
                          \epsilon^3 \Bigg( \frac{6333061}{34992} -
                          \frac{11531}{648} \zeta_2  + \frac{1499}{240}
                          \zeta_2^2 
                          \nonumber\\
                        &+ 
                          \frac{253}{1680} \zeta_2^3 +
                          \frac{19415}{972} \zeta_3  - \frac{235}{36}
                          \zeta_2 \zeta_3  - 
                          \frac{1153}{108} \zeta_3^2 + \frac{535}{36}
                          \zeta_5 \Bigg)  
                          + 
                          \epsilon^4 \Bigg( - \frac{128493871}{419904}
                          \nonumber\\
                        &+ \frac{133237}{3888} \zeta_2  -
                          \frac{21533}{2592} \zeta_2^2  + 
                          \frac{649}{1440} \zeta_2^3 -
                          \frac{156127}{5832} \zeta_3  + \frac{215}{27}
                          \zeta_2 \zeta_3  + 
                          \frac{517}{80} \zeta_2^2 \zeta_3 +
                          \frac{14675}{648} \zeta_3^2  
                          \nonumber\\
                        &-
                          \frac{2204}{135} \zeta_5  + 
                          \frac{171}{40} \zeta_2 \zeta_5 +
                          \frac{229}{336} \zeta_7 \Bigg) \Bigg\}  
                          + 
                          {\dis{C_{A}^2}} \Bigg\{ 
                          \frac{32}{\epsilon^4}  + \frac{44}{3
                          \epsilon^3} + \Bigg( - \frac{422}{9}  - 4
                          \zeta_2 \Bigg) \frac{1}{\epsilon^2}
                          +  
                          \Bigg( \frac{890}{27} 
                          \nonumber\\
                        &- 11 \zeta_2 +
                          \frac{50}{3} \zeta_3 \Bigg)
                          \frac{1}{\epsilon} + \frac{3835}{81} +  
                          \frac{115}{6} \zeta_2 - \frac{21}{5}
                          \zeta_2^2 + \frac{11}{9} \zeta_3  
                          + 
                          \epsilon \Bigg( - \frac{213817}{972} -
                          \frac{103}{18} \zeta_2  + \frac{77}{120}
                          \zeta_2^2 
                          \nonumber\\
                        &+ \frac{1103}{54} \zeta_3  - 
                          \frac{23}{6} \zeta_2 \zeta_3 - \frac{71}{10}
                          \zeta_5 \Bigg)  
                          + 
                          \epsilon^2 \Bigg( \frac{6102745}{11664} -
                          \frac{991}{27} \zeta_2  - \frac{2183}{240} \zeta_2^2 + 
                          \frac{2313}{280} \zeta_2^3 - \frac{8836}{81}
                          \zeta_3  
                          \nonumber\\
                        &- \frac{55}{12} \zeta_2 \zeta_3 + 
                          \frac{901}{36} \zeta_3^2 + \frac{341}{60}
                          \zeta_5 \Bigg)  
                          + 
                          \epsilon^3 \Bigg( - \frac{142142401}{139968}
                          + \frac{75881}{648} \zeta_2  +
                          \frac{79819}{2160} \zeta_2^2 -  
                          \frac{2057}{480} \zeta_2^3 
                          \nonumber\\
                        &+
                          \frac{606035}{1944} \zeta_3 -  \frac{251}{72}
                          \zeta_2 \zeta_3 -  
                          \frac{1291}{80} \zeta_2^2 \zeta_3 -
                          \frac{5137}{216} \zeta_3^2 +
                          \frac{14459}{360} \zeta_5 +  
                          \frac{313}{40} \zeta_2 \zeta_5 -
                          \frac{3169}{28} \zeta_7 \Bigg) 
                          \nonumber\\
                        &+  
                          \epsilon^4 \Bigg( \frac{2999987401}{1679616}
                          - \frac{1943429}{7776} \zeta_2  -
                          \frac{15707}{160} \zeta_2^2 -  
                          \frac{35177}{20160} \zeta_2^3 +
                          \frac{50419}{1600} \zeta_2^4  -
                          \frac{16593479}{23328} \zeta_3 
                          \nonumber\\
                        &+  
                          \frac{1169}{27} \zeta_2 \zeta_3 +
                          \frac{22781}{1440} \zeta_2^2 \zeta_3 +  
                          \frac{93731}{1296} \zeta_3^2 -
                          \frac{1547}{144} \zeta_2 \zeta_3^2 -
                          \frac{8137}{54} \zeta_5 -  
                          \frac{1001}{80} \zeta_2 \zeta_5 +
                          \frac{845}{24} \zeta_3 \zeta_5 
                          \nonumber\\
                        &-
                          \frac{33}{2} \zeta_{5,3} +  
                          \frac{56155}{672} \zeta_7 \Bigg) \Bigg\}\,,
                          \nonumber\\ 
  {\hat \F}^{A,(3)}_{g} &=  {\dis{n_{f} C^{(2)}_{J}}} \Bigg\{ - 2+3 \epsilon \Bigg\}
                          + 
                          {\dis{C_{F} n_{f}^2}} \Bigg\{ \Bigg( -
                          \frac{640}{9} + 16 \ln
                          \left(\frac{q^2}{m_t^2}\right) + \frac{64}{3}
                          \zeta_3  \Bigg) \frac{1}{\epsilon} +  
                          \frac{7901}{27}  
                          \nonumber\\
                        &- 24 \ln
                          \left(\frac{q^2}{m_t^2}\right) 
                          - \frac{32}{3}
                          \zeta_2  - \frac{112}{15} \zeta_2^2 - 
                          \frac{848}{9} \zeta_3 \Bigg\} 
                          + 
                          {\dis{C_{F}^2 n_{f}}} \Bigg\{ \frac{457}{6} + 104
                          \zeta_3  - 160 \zeta_5 \Bigg\} 
                          \nonumber\\
                        &+ 
                          {\dis{C_{A}^2 n_{f}}} \Bigg\{ 
                          \frac{64}{3 \epsilon^5} - \frac{32}{81
                          \epsilon^4}  
                          + 
                          \Bigg( - \frac{18752}{243}  - \frac{376}{27}
                          \zeta_2 \Bigg)  \frac{1}{\epsilon^3}  + 
                          \Bigg( \frac{36416}{243} - \frac{1700}{81}
                          \zeta_2  + \frac{2072}{27} \zeta_3 \Bigg)
                          \frac{1}{\epsilon^2} 
                          \nonumber\\
                        &+ \Bigg(
                          \frac{62642}{2187}  + \frac{22088}{243}
                          \zeta_2 
                          - \frac{2453}{90} \zeta_2^2  - 
                          \frac{3988}{81} \zeta_3 \Bigg)
                          \frac{1}{\epsilon} -
                          \frac{14655809}{13122} -
                          \frac{60548}{729} \zeta_2  + 
                          \frac{917}{60} \zeta_2^2 
                          \nonumber\\
                        &- \frac{772}{27} \zeta_3
                          - \frac{439}{9} \zeta_2 \zeta_3   
                          + \frac{3238}{45}
                          \zeta_5 \Bigg\}  
                          + 
                          {\dis{C_{A}  n_{f}^2}} \Bigg\{ - \frac{128}{81
                          \epsilon^4} + \frac{640}{243 \epsilon^3}  + \Bigg(
                          \frac{128}{27}  + \frac{80}{27} \zeta_2
                          \Bigg) \frac{1}{\epsilon^2}  
                          \nonumber\\
                        &+ 
                          \Bigg( - \frac{93088}{2187} 
                          - \frac{400}{81}
                          \zeta_2  
                          - \frac{1328}{81} \zeta_3 \Bigg)
                          \frac{1}{\epsilon} + 
                          \frac{1066349}{6561}  -
                          \frac{56}{27} \zeta_2  + 
                          \frac{797}{135} \zeta_2^2  + 
                          \frac{13768}{243} \zeta_3 \Bigg\} 
                          \nonumber\\
                        &+ 
                          {\dis{C_{A} C_{F} 
                          n_{f}}} \Bigg\{ -
                          \frac{16}{9 \epsilon^3} 
                          +  
                          \Bigg( \frac{5980}{27} - 48 \ln \left(\frac{q^2}{m_t^2}\right)  -
                          \frac{640}{9} \zeta_3 \Bigg)
                          \frac{1}{\epsilon^2} + \Bigg( -
                          \frac{20377}{81} 
                          \nonumber\\
                        &- 16 \ln \left(\frac{q^2}{m_t^2}\right)  + \frac{86}{3}
                          \zeta_2  + 
                          \frac{352}{15} \zeta_2^2 
                          + \frac{1744}{27}
                          \zeta_3 \Bigg) \frac{1}{\epsilon} + 72 \ln \left(\frac{q^2}{m_t^2}\right)   
                          - \frac{587705}{972} - 
                          \frac{551}{6} \zeta_2 
                          \nonumber\\
                        &+ 12 \ln \left(\frac{q^2}{m_t^2}\right) \zeta_2 -
                          \frac{96}{5} \zeta_2^2  + \frac{12386}{81}
                          \zeta_3  
                          + 
                          48 \zeta_2 \zeta_3  +
                          \frac{32}{9} \zeta_5 \Bigg\}   
                          + 
                          {\dis{C_{A}^3}} \Bigg\{ -
                          \frac{256}{3 \epsilon^6}  - \frac{352}{3
                          \epsilon^5}  
                          \nonumber\\
                        &+ \frac{16144}{81 \epsilon^4}  + 
                          \Bigg( \frac{22864}{243} + \frac{2068}{27}
                          \zeta_2  - \frac{176}{3} \zeta_3 \Bigg)
                          \frac{1}{\epsilon^3}  
                          + 
                          \Bigg( - \frac{172844}{243} - \frac{1630}{81}
                          \zeta_2  + \frac{494}{45} \zeta_2^2  
                          \nonumber\\
                        &-
                          \frac{836}{27} \zeta_3 \Bigg)
                          \frac{1}{\epsilon^2}  + \Bigg(
                          \frac{2327399}{2187}  - \frac{71438}{243}
                          \zeta_2 +  
                          \frac{3751}{180} \zeta_2^2 
                          - \frac{842}{9}
                          \zeta_3  + \frac{170}{9} \zeta_2 \zeta_3  + 
                          \frac{1756}{15} \zeta_5 \Bigg)
                          \frac{1}{\epsilon} 
                          \nonumber\\
                        &+ \frac{16531853}{26244} + 
                          \frac{918931}{1458} \zeta_2 +
                          \frac{27251}{1080} \zeta_2^2  -
                          \frac{22523}{270} \zeta_2^3   
                          - \frac{51580}{243}
                          \zeta_3  + \frac{77}{18} \zeta_2 \zeta_3 -
                          \frac{1766}{9} \zeta_3^2  
                          \nonumber\\
                        &+ 
                          \frac{20911}{45} \zeta_5 \Bigg\}\,.
\end{align}
The results up to two loop level is consistent with the existing
ones~\cite{Ravindran:2004mb} and the three loop result is the new
one. These are required in the context of computing SV cross-section
which is discussed below.

The form factor ${\F}^{A}_{g}(\hat{a}_{s}, Q^{2}, \mu^{2}, \epsilon)$
satisfies the $KG$-differential equation \cite{Sudakov:1954sw,
  Mueller:1979ih, Collins:1980ih, Sen:1981sd, Magnea:1990zb} which is
a direct consequence of the facts that QCD amplitudes exhibit
factorisation property, gauge and renormalisation group (RG)
invariances:
\begin{equation}
  \label{eq:KG}
  Q^2 \frac{d}{dQ^2} \ln {\F}^{A}_{g} (\hat{a}_s, Q^2, \mu^2, \epsilon)
  = \frac{1}{2} \left[ K^{A}_{g} \left(\hat{a}_s, \frac{\mu_R^2}{\mu^2}, \epsilon
    \right)  + G^{A}_{g} \left(\hat{a}_s, \frac{Q^2}{\mu_R^2}, \frac{\mu_R^2}{\mu^2}, \epsilon \right) \right]\,.
\end{equation}
In the above expression, all the poles in dimensional regularisation
parameter $\ep$ are captured in the $Q^{2}$ independent function
$K^{A}_{g}$ and the quantities which are finite as
$\epsilon \rightarrow 0$ are encapsulated in $G^{A}_{g}$. The
solutions of the $KG$ equation in the desired form is given
in~\cite{Ravindran:2005vv} as (see also
\cite{Ahmed:2014cla,Ahmed:2014cha})
\begin{align}
  \label{eq:lnFSoln}
  \ln {\F}^{A}_{g}(\hat{a}_s, Q^2, \mu^2, \ep) =
  \sum_{i=1}^{\infty} {\hat a}_{s}^{i} \l(\frac{Q^{2}}{\mu^{2}}\r)^{i
  \frac{\ep}{2}} S_{\ep}^{i} \hat {\cal L}_{g,i}^{A}(\ep)
\end{align}
with
\begin{align}
  \label{eq:lnFitoCalLF}
  \hat {\cal L}_{g,1}^{A}(\ep) =& { \frac{1}{\ep^2} } \Bigg\{-2 A^{A}_{g,1}\Bigg\}
                                  + { \frac{1}{\ep}
                                  }
                                  \Bigg\{G^{A}_{g,1}
                                  (\ep)\Bigg\}\, ,
                                  \nonumber\\
  \hat {\cal L}_{g,2}^{A}(\ep) =& { \frac{1}{\ep^3} } \Bigg\{\beta_0 A^{A}_{g,1}\Bigg\}
                                  + {
                                  \frac{1}{\ep^2} }
                                  \Bigg\{-  {
                                  \frac{1}{2} }  A^{A}_{g,2}
                                  - \beta_0   G^{A}_{g,1}(\ep)\Bigg\}
                                  + { \frac{1}{\ep}
                                  } \Bigg\{ {
                                  \frac{1}{2} }  G^{A}_{g,2}(\ep)\Bigg\}\, ,
                                  \nonumber\\
  \hat {\cal L}_{g,3}^{A}(\ep) =& { \frac{1}{\ep^4}
                                  } \Bigg\{- {
                                  \frac{8}{9} }  \beta_0^2 A^{A}_{g,1}\Bigg\}
                                  + {
                                  \frac{1}{\ep^3} }
                                  \Bigg\{ { \frac{2}{9} } \beta_1 A^{A}_{g,1}
                                  + { \frac{8}{9} }
                                  \beta_0 A^{A}_{g,2}  + { \frac{4}{3} }
                                  \beta_0^2 G^{A}_{g,1}(\ep)\Bigg\}
                                  \nonumber\\
                                &
                                  + { \frac{1}{\ep^2} } \Bigg\{- { \frac{2}{9} } A^{A}_{g,3}
                                  - { \frac{1}{3} } \beta_1 G^{A}_{g,1}(\ep)
                                  - { \frac{4}{3} } \beta_0 G^{A}_{g,2}(\ep)\Bigg\}
                                  + { \frac{1}{\ep}
                                  } \Bigg\{  { \frac{1}{3} } G^{A}_{g,3}(\ep)\Bigg\}\, .
\end{align}
$A^{A}_{g}$'s are called the cusp anomalous dimensions.  The constants
$G^{A}_{g,i}$'s are the coefficients of $a_{s}^{i}$ in the following
expansions:
\begin{align}
  \label{eq:GandAExp}
  G^{A}_{g}\left(\hat{a}_s, \frac{Q^2}{\mu_R^2}, \frac{\mu_R^2}{\mu^2},
  \epsilon \right) &= G^{A}_{g}\left(a_{s}(Q^{2}), 1, \epsilon \right)
                     + \int_{\frac{Q^{2}}{\mu_{R}^{2}}}^{1}
                     \frac{dx}{x} A^{A}_{g}(a_{s}(x\mu_{R}^{2}))
                     \nonumber\\
                   &= \sum_{i=1}^{\infty}a_{s}^{i}(Q^{2}) G^{A}_{g,i}(\epsilon) + \int_{\frac{Q^{2}}{\mu_{R}^{2}}}^{1}
                     \frac{dx}{x} A^{A}_{g}(a_{s}(x\mu_{R}^{2}))\,.
\end{align}
However, the solutions of the logarithm of the form factor involves
the unknown functions $G^{A}_{g,i}$ which are observed to fulfil
\cite{Ravindran:2004mb, Moch:2005tm} the following decomposition in
terms of collinear ($B^{A}_{g}$), soft ($f^{A}_{g}$) and UV
($\gamma^{A}_{g}$) anomalous dimensions:
\begin{align}
  \label{eq:GIi}
  G^{A}_{g,i} (\ep) = 2 \left(B^{A}_{g,i} -
  \gamma^{A}_{g,i}\right)  + f^{A}_{g,i} +
  C^{A}_{g,i}  + \sum_{k=1}^{\infty} \epsilon^k g^{A,k}_{g,i} \, ,
\end{align}
where, the constants $C^{A}_{g,i}$ are given by
\cite{Ravindran:2006cg}
\begin{align}
  \label{eq:Cg}
  C^{A}_{g,1} &= 0\, ,
                \nonumber\\
  C^{A}_{g,2} &= - 2 \beta_{0} g^{A,1}_{g,1}\, ,
                \nonumber\\
  C^{A}_{g,3} &= - 2 \beta_{1} g^{A,1}_{g,1} - 2
                \beta_{0} \left(g^{A,1}_{g,2}  + 2 \beta_{0} g^{A,2}_{g,1}\right)\, .
\end{align}
In the above expressions, $X^{A}_{g,i}$ with $X=A,B,f$ and
$\gamma^{A}_{g, i}$ are defined through the series expansion in powers
of $a_{s}$:
\begin{align}
  \label{eq:ABfgmExp}
  X^{A}_{g} &\equiv \sum_{i=1}^{\infty} a_{s}^{i}
              X^{A}_{g,i}\,,
              \qquad \text{and} \qquad
              \gamma^{A}_{g} \equiv \sum_{i=1}^{\infty} a_{s}^{i} \gamma^{A}_{g,i}\,\,.
\end{align}
$f_{g}^{A}$'s are introduced for the first time in the
article~\cite{Ravindran:2004mb} where it is shown to fulfil the
maximally non-Abelian property up to two loop level whose validity is
reconfirmed in~\cite{Moch:2005tm} at three loop level. Moreover, due
to universality of the quantities denoted by $X$, these are
independent of the operator insertion. These are only dependent on the
initial state partons of any process. Hence, being a process of gluon
fusion, we can make use of the existing results up to three loop:
\begin{align}
  \label{eq:1}
  X^{A}_{g} = X_{g}\,.
\end{align}
$f_{g}$ can be found in \cite{Ravindran:2004mb, Moch:2005tm},
$A_{g,i}$ in~\cite{Moch:2004pa, Moch:2005tm, Vogt:2004mw, Vogt:2000ci}
and $B_{g,i}$ in \cite{Vogt:2004mw, Moch:2005tm} up to three loop
level. Utilising the results of these known quantities and comparing
the above expansion of $G^{A}_{g,i}(\ep)$, Eq.~(\ref{eq:GIi}), with
the results of the logarithm of the form factors, we extract the
relevant $g_{g,i}^{A,k}$ and $\gamma^{A}_{g,i}$'s up to three
loop. For soft-virtual cross section at N$^{3}$LO we need
$g^{A,1}_{g.3}$ in addition to the quantities arising from one and two
loop. The form factors for the pseudo-scalar production up to two loop
can be found in~\cite{Ravindran:2004mb} and the three loop one is
calculated very recently in the article~\cite{Ahmed:2015qpa} by some
of us. However, in this computation of SV cross section at N$^{3}$LO,
we need the form factor in a particular form which is little bit
different than the ones presented in our recent
article~\cite{Ahmed:2015qpa}, though the required one can be extracted
from the results provided there. For readers' convenience, we have
presented the form factors $\F^{A}_{g}$ up to three loop in the
beginning of this section which have been employed to extract the
required $g^{A,k}_{g,i}$'s using Eq.~(\ref{eq:lnFSoln}),
(\ref{eq:lnFitoCalLF}) and (\ref{eq:GIi}). Below we present our
finding of the relevant $g^{A,k}_{g,i}$'s up to three loop level:
\begin{align}
  \label{eq:g31}
  g^{A,1}_{g,1} &= {\dis{C_{A}}} \Bigg\{ 4 + \zeta_2 \Bigg\}\,,
                  \nonumber\\
  g^{A,2}_{g,1} &= {\dis{C_{A}}} \Bigg\{ - 6 - \frac{7}{3} \zeta_3 \Bigg\}\,,
                  \nonumber\\
  g^{A,3}_{g,1} &= {\dis{C_{A}}} \Bigg\{ 7 - \frac{1}{2} \zeta_2 + \frac{47}{80} \zeta_2^2 \Bigg\}\,,
                  \nonumber\\ \nonumber\\
  g^{A,1}_{g,2} &= {\dis{C_{A}^2}} \Bigg\{ \frac{11882}{81} + \frac{67}{3}
                  \zeta_2 - \frac{44}{3} \zeta_3 \Bigg\} 
                  + 
                  {\dis{C_{A} n_{f}}} \Bigg\{ - \frac{2534}{81} - \frac{10}{3} \zeta_2 -
                  \frac{40}{3} \zeta_3 \Bigg\} 
                  + 
                  {\dis{C_{F} n_{f}}} \Bigg\{ - \frac{160}{3} 
                  \nonumber\\
                &+ 12 \ln \left(\frac{\mu_R^2}{m_t^2}\right) + 16 \zeta_3 \Bigg\}\,,
                  \nonumber\\
  g^{A,2}_{g,2} &= {\dis{C_{F} n_{f}}} \Bigg\{ \frac{2827}{18} - 18
                  \ln \left(\frac{\mu_R^2}{m_t^2}\right)  - \frac{19}{3} \zeta_2 - \frac{16}{3}
                  \zeta_2^2  - 
                  \frac{128}{3} \zeta_3 \Bigg\} 
                  + {\dis{C_{A} n_{f}}} \Bigg\{
                  \frac{21839}{243}  - \frac{17}{9} \zeta_2 
                  \nonumber\\
                &+
                  \frac{259}{60} \zeta_2^2  + 
                  \frac{766}{27} \zeta_3 \Bigg\} 
                  + {\dis{C_{A}^2}} \Bigg\{ -
                  \frac{223861}{486}  + \frac{80}{9} \zeta_2 +
                  \frac{671}{120} \zeta_2^2  + 
                  \frac{2111}{27} \zeta_3 + \frac{5}{3} \zeta_2
                  \zeta_3  - 39 \zeta_5 \Bigg\}\,,
                  \nonumber\\ \nonumber\\
  g^{A,1}_{g,3} &=  {\dis{n_{f} C_{J}^{(2)}}}  \Bigg\{ - 6 \Bigg\}  
                  + {\dis{C_{F}
                  n_{f}^2}} \Bigg\{ \frac{12395}{27}  -
                  \frac{136}{9} \zeta_2  - 
                  \frac{368}{45} \zeta_2^2 - \frac{1520}{9} \zeta_3  -
                  24  \ln \left(\frac{\mu_R^2}{m_t^2}\right) 
                  \Bigg\}  
                  \nonumber\\
                &+ 
                  {\dis{C_{F}^2 n_{f}}} \Bigg\{ \frac{457}{2} + 312 \zeta_3 -
                  480 \zeta_5 \Bigg\}  
                  + 
                  {\dis{C_{A}^2 n_{f}}} \Bigg\{ - \frac{12480497}{4374} -
                  \frac{2075}{243} \zeta_2  - \frac{128}{45} \zeta_2^2 
                  \nonumber\\
                &- 
                  \frac{12992}{81} \zeta_3 - \frac{88}{9} \zeta_2
                  \zeta_3 + \frac{272}{3} \zeta_5 \Bigg\}
                  + 
                  {\dis{C_{A}^3}} \Bigg\{ \frac{62867783}{8748} +
                  \frac{146677}{486} \zeta_2  - \frac{5744}{45}
                  \zeta_2^2  - 
                  \frac{12352}{315} \zeta_2^3 
                  \nonumber\\
                &- \frac{67766}{27}
                  \zeta_3 - \frac{1496}{9} \zeta_2 \zeta_3  - 
                  \frac{104}{3} \zeta_3^2 + \frac{3080}{3} \zeta_5 \Bigg\}
                  + 
                  {\dis{C_{A} n_{f}^2}} \Bigg\{ \frac{514997}{2187} -
                  \frac{8}{27} \zeta_2  + \frac{232}{45} \zeta_2^2 
                  \nonumber\\
                &+ 
                  \frac{7640}{81} \zeta_3 \Bigg\} 
                  + {\dis{C_{A} C_{F} n_{f}}} \Bigg\{
                  - \frac{1004195}{324}  + \frac{1031}{18} \zeta_2 + 
                  \frac{1568}{45} \zeta_2^2 + \frac{25784}{27} \zeta_3
                  + 40 \zeta_2 \zeta_3  + \frac{608}{3} \zeta_5 
                  \nonumber\\
                &+ 132 \ln \left(\frac{\mu_R^2}{m_t^2}\right) \Bigg\}\,.
\end{align}
The component of the Wilson coefficient, $C^{(2)}_{J}$, which is
defined through Eq.~(\ref{eq:const}), is not available in the
literature. The other constants $\gamma^{A}_{g,i}$ up to three loop
($i=3$) are obtained as
\begin{align}
  \label{eq:gamma3}
  \gamma^{A}_{g,1} &= \frac{11}{3} C_{A} - \frac{2}{3} n_{f}\,, 
                     \nonumber\\
  \gamma^{A}_{g,2} &=  \frac{34}{3} C_{A}^{2} - \frac{10}{3} C_{A} n_{f} -
                     2 C_{F} n_{f}\,,
                     \nonumber\\
  \gamma^{A}_{g,3} &=  \frac{2857}{54} C_{A}^3 - \frac{1415}{54} C_{A}^2
                     n_{f}  
                     - \frac{205}{18} C_{A} C_{F} n_{f} + C_{F}^2 n_{f} +
                     \frac{79}{54} C_{A} n_{f}^2 + \frac{11}{9} C_{F} n_{f}^2\,.
\end{align}
As a matter of emphasising the fact, note that the
$\gamma^{A}_{g,i}$'s are found to satisfy
\begin{align}
  \label{eq:gammaBeta}
  \gamma^{A}_{g} = -\frac{\beta}{a_{s}} \qquad \text{up to 3-loop}\,,
\end{align}
where, $\beta = - \sum_{i=0}^{\infty} \beta_{i} a_{s}^{i+2}$ is the
usual QCD $\beta$-function. For more elaborate discussion on this, see
recent article~\cite{Ahmed:2015qpa} (also see \cite{Larin:1993tq,
  Zoller:2013ixa}).

\subsection{Operator Renormalisation Constant}
\label{ss:OOR}

The strong coupling constant renormalisation through $Z_{a_{s}}$ is
not sufficient to make the form factor $\F^{A}_{g}$ completely UV
finite, one needs to perform additional renormalisation to remove the
residual UV divergences which is reflected through the presence of
non-zero $\gamma^{A}_{g}$ in Eq.~(\ref{eq:GIi}). This additional
renormalisation is called the overall operator renormalisation which
is performed through the constant $Z^{A}_{g}$. This is determined by
solving the underlying RG equation:
\begin{align}
  \label{eq:ZRGE}
  \mu_{R}^{2} \frac{d}{d\mu_{R}^{2}} \ln Z^{A}_{g} \left( {\hat a}_{s},
  \mu_{R}^{2}, \mu^{2}, \epsilon \right) = \sum_{i=1}^{\infty}
  a_{s}^{i} \gamma^{A}_{g,i}\,. 
\end{align}
Using the results of $\gamma^{A}_{g,i}$ from Eq.~(\ref{eq:gamma3}) and
solving the above RG equation, we obtain the overall renormalisation
constant up to three loop level given by
\begin{align}
  \label{eq:ZGG}
  Z^{A}_{g} &= 1 +  a_s \Bigg[ \frac{22}{3\epsilon}
              C_{A}  -
              \frac{4}{3\epsilon} n_{f} \Bigg] 
              + 
              a_s^2 \Bigg[ \frac{1}{\epsilon^2}
              \Bigg\{ \frac{484}{9} C_{A}^2 - \frac{176}{9} C_{A}
              n_{f} + \frac{16}{9} n_{f}^2 \Bigg\}
              + \frac{1}{\epsilon} \Bigg\{ \frac{34}{3} C_{A}^2  
              \nonumber\\
            &-
              \frac{10}{3} C_{A} n_{f}  - 2 C_{F} n_{f} \Bigg\} \Bigg] 
              + 
              a_s^3 \Bigg[   \frac{1}{\epsilon^3} 
              \Bigg\{ \frac{10648}{27} C_{A}^3 - \frac{1936}{9}
              C_{A}^2 n_{f}  + \frac{352}{9} C_{A} n_{f}^2  -
              \frac{64}{27} n_{f}^3 \Bigg\}  
              \nonumber\\
            &+   \frac{1}{\epsilon^2}
              \Bigg\{ \frac{5236}{27} C_{A}^3 - \frac{2492}{27}
              C_{A}^2 n_{f}  - \frac{308}{9} C_{A} C_{F} n_{f}  + 
              \frac{280}{27} C_{A} n_{f}^2  + \frac{56}{9} C_{F}
              n_{f}^2 \Bigg\}
              \nonumber\\
            &  
              +  \frac{1}{\epsilon} \Bigg\{ \frac{2857}{81} C_{A}^3  -
              \frac{1415}{81} C_{A}^2 n_{f}  - \frac{205}{27} C_{A} C_{F} n_{f} + 
              \frac{2}{3} C_{F}^2 n_{f} + \frac{79}{81} C_{A}
              n_{f}^2  + \frac{22}{27} C_{F} n_{f}^2 \Bigg\}
              \Bigg] \, .
\end{align}
We emphasise that $Z_{g}^{A}=Z_{GG}$ which is introduced in
Eq.~(\ref{eq:FFDefMatEle}) has been discussed in great detail
in~\cite{Ahmed:2015qpa}.  The complete UV finite form factor
$[\F^{A}_{g}]_{R}$ in terms of this $Z_{g}^{A}$ is
\begin{align}
  \label{eq:FFRen}
  [\F^{A}_{g}]_{R} = Z^{A}_{g} \F^{A}_{g}\,.
\end{align}
This is presented in our recent article~\cite{Ahmed:2015qpa} up to
three loops in the form of hard matching coefficients of
soft-collinear effective theory.

\subsection{Mass Factorisation Kernel}
\label{ss:MFK}

The UV finite form factor contains additional divergences arising from
the soft and collinear regions of the loop momenta. In this section,
we address the issue of collinear divergences and describe a
prescription to remove them. The collinear singularities that arise in
the massless limit of partons are removed in the ${\overline {MS}}$
scheme using mass factorisation kernel
$\Gamma(\hat{a}_s, \mu^2, \mu_F^2, z, \epsilon)$. The kernel satisfies
the following RG equation :
\begin{align}
  \label{eq:kernelRGE}
  \mu_F^2 \frac{d}{d\mu_F^2} \Gamma(z,\mu_F^2,\epsilon) = \frac{1}{2} P \left(z,\mu_F^2\right) \otimes \Gamma \left(z,\mu_F^2,\epsilon \right) 
\end{align}
where, $P\left(z,\mu_{F}^{2}\right)$ are Altarelli-Parisi splitting
functions (matrix valued). Expanding $P\left(z,\mu_{F}^{2}\right)$ and
$\Gamma(z,\mu_F^2,\epsilon)$ in powers of the strong coupling constant
we get
\begin{align}
  \label{eq:APexpand}
  &P(z,\mu_{F}^{2}) = \sum_{i=1}^{\infty} a_{s}^{i}(\mu_{F}^{2})P^{(i-1)}(z)\, 
    \intertext{and}
  &\Gamma(z,\mu_F^2,\epsilon) = \delta(1-z) + \sum_{i=1}^{\infty}
    {\hat a}_{s}^{i}  \l(\frac{\mu_{F}^{2}}{\mu^{2}}\r)^{i
    \frac{\ep}{2}} S_{\ep}^{i}  \Gamma^{(i)}(z,\ep)\, .
\end{align}
The RG equation of $\Gamma(z,\mu_F^2,\epsilon)$,
Eq.~(\ref{eq:kernelRGE}), can be solved in dimensional regularisation
in powers of ${\hat a}_{s}$.  In the $\overline{MS}$ scheme, the
kernel contains only the poles in $\ep$. The solutions up to the
required order $\Gamma^{(3)}(z,\epsilon)$ in terms of $P^{(i)}(z)$ can
be found in Eq.~(33) of \cite{Ravindran:2005vv}. The relevant ones up
to three loop, $P^{(0)}(z), P^{(1)}(z) ~\text{and}~ P^{(2)}(z)$ are
computed in the articles~\cite{Moch:2004pa, Vogt:2004mw}. For the SV
cross section only the diagonal parts of the splitting functions
$P^{(i)}_{gg}(z)$ and kernels $\Gamma^{(i)}_{gg}(z,\ep)$ contribute.

\subsection{Soft-Collinear Distribution}
\label{ss:SCD}

The resulting expression from form factor along with operator
renormalisation constant and mass factorisation kernel is not
completely finite, it contains some residual divergences which get
cancelled against the contribution arising from soft gluon
emissions. Hence, the finiteness of $\Delta_{g}^{A, \sv}$ in the limit
$\ep \rightarrow 0$ demands that the soft-collinear distribution,
$\Phi^A_{g} (\hat{a}_s, q^2, \mu^2, z, \epsilon)$, has pole structure
in $\ep$ similar to that of residual divergences. In
articles~\cite{Ravindran:2005vv} and \cite{Ravindran:2006cg} it was
shown that $\Phi^{A}_{g}$ must obey $KG$ type integro-differential
equation, which we call ${\overline{KG}}$ equation, to remove that
residual divergences:
\begin{equation}
  \label{eq:KGbarEqn}
  q^2 \frac{d}{dq^2} \Phi^A_{g}\left(\hat{a}_s, q^2, \mu^2, z,
    \ep\right)   = \frac{1}{2} \left[ \overline K^A_{g} \left(\hat{a}_s, \frac{\mu_R^2}{\mu^2}, z,
      \ep \right)  + \overline G^A_{g} \left(\hat{a}_s,
      \frac{q^2}{\mu_R^2},  \frac{\mu_R^2}{\mu^2}, z, \ep \right) \right] \, .
\end{equation}
${\overline K}$ and ${\overline G}$ play similar roles as those of $K$
and $G$, respectively. Also,
$\Phi^A_{g} (\hat{a}_s, q^2, \mu^2, z, \ep)$ being independent of
$\mu_{R}^{2}$ satisfy the RG equation
\begin{align}
  \label{eq:RGEphi}
  \mu_{R}^{2}\frac{d}{d\mu_{R}^{2}}\Phi^A_{g} (\hat{a}_s, q^2, \mu^2, z, \epsilon) = 0\, .
\end{align}
This RG invariance and the demand of cancellation of all the residual
divergences arising from $\F^{A}_{g}, Z^{A}_{g}$ and $\Gamma_{gg}$
against $\Phi^{A}_{g}$ implies the solution of the ${\overline {KG}}$
equation as~\cite{Ravindran:2005vv, Ravindran:2006cg}
\begin{align}
  \label{eq:PhiSoln}
  \Phi^A_{g} (\hat{a}_s, q^2, \mu^2, z, \epsilon) &= \Phi^A_{g} (\hat{a}_s, q^2(1-z)^{2}, \mu^2, \epsilon)
                                                    \nn\\
                                                  &= \sum_{i=1}^{\infty}
                                                    {\hat a}_{s}^{i}  
                                                    \l(\frac{q^{2}(1-z)^{2}}{\mu^{2}}\r)^{i
                                                    \frac{\ep}{2}}  S_{\ep}^{i}  
                                                    \l(\frac{i\ep}{1-z}\r) {\hat \phi}^{A}_{g,i}(\ep)
\end{align}
with
\begin{align}
  \label{eq:phiHatIi}
  {\hat \phi}^{A}_{g,i}(\ep) = {\cal L}_{g,i}^{A}(\ep)\l(A^{A}_{g,j}
  \rightarrow - A^{A}_{g,j},  G^{A}_{g,j}(\ep) \rightarrow {\overline{\cal G}}^{A}_{g,j}(\ep)\r)
\end{align}
where, ${\cal L}_{g,i}^{A}(\ep)$ are defined in
Eq.~(\ref{eq:lnFitoCalLF}).
The z-independent constants ${\overline{\cal G}}^{A}_{g,i}(\ep)$ can
be obtained by comparing the poles as well as non-pole terms in $\ep$
of ${\hat \phi}^{A}_{g,i}(\ep)$ with those arising from form factor,
overall renormalisation constant and splitting functions. We find
\begin{align}
  \label{eq:calGexpans}
  \overline {\cal G}^{A}_{g,i}(\ep)&= - f_{g,i}^A + {\overline
                                     C}_{g,i}^{A}  + \sum_{k=1}^\infty \ep^k {\overline {\cal G}}^{A,k}_{g,i} \, ,
\end{align}
where,
\begin{align}
  \label{eq:overlineCiI}
  &{\overline C}_{g,1}^{A} = 0\, ,
    \nonumber\\
  &{\overline C}_{g,2}^{A} = -2\beta_{0}{\g}_{g,1}^{A,1}\, ,
    \nonumber\\
  &{\overline C}_{g,3}^{A} = -2\beta_{1}{\g}_{g,1}^{A,1} -
    2\beta_{0}\left({\g}_{g,2}^{A,1}  + 2\beta_{0}{\g}_{g,1}^{A,2} \right)\, .
\end{align}
However, due to the universality of the soft gluon contribution,
$\Phi^{A}_{g}$ must be the same as that of the Higgs boson production
in gluon fusion:
\begin{align}
  \label{eq:PhiAgPhiHg}
  \Phi^{A}_{g} &= \Phi^{H}_{g} = \Phi_{g}
                 \nonumber\\
  \text{i.e.}~~{\g}^{A,k}_{g,i} &= {\g}^{H,k}_{g,i} = {\g}^{k}_{g,i}\,. 
\end{align}
In the above expression, $\Phi_{g}$ and ${\g}^{k}_{g,i}$ are written
in order to emphasise the universality of these quantities i.e.
$\Phi^{H}_{g}$ and ${\g}^{H,k}_{g,i}$ can be used for any gluon fusion
process, these are independent of the operator insertion. The relevant
constants ${\g}_{g,1}^{H,1},{\g}_{g,1}^{H,2},{\g}_{g,2}^{H,1}$ are
determined from the result of the explicit computations of soft gluon
emission to the Higgs boson production \cite{Ravindran:2003um}. Later,
these corrections are extended to all orders in dimensional
regularisation parameter $\ep$ in the article \cite{deFlorian:2012za},
using which we extract ${\g}_{g,1}^{H,3}$ and ${\g}_{g,2}^{H,2}$.  The
third order constant ${\g}_{g,3}^{H,1}$ is computed from the result of
SV cross section for the production of the Higgs boson at N$^{3}$LO
\cite{Anastasiou:2014vaa}. This was presented in the
article~\cite{Ahmed:2014cla}. The ${\g}_{g,i}^{H,k}$'s required to get
the SV cross sections up to N$^{3}$LO are listed below:
\begin{align}
  \label{eq:calG}
  {\overline {\cal G}}^{H,1}_{g,1} &= {\dis{C_A}} \Bigg\{ - 3 \zeta_2 \Bigg\} \,, 
                                     \nonumber\\
  {\overline {\cal G}}^{H,2}_{g,1} &= {\dis{C_A}} \Bigg\{ \frac{7}{3} \zeta_3 \Bigg\} \,, 
                                     \nonumber\\
  {\overline {\cal G}}^{H,3}_{g,1} &=  {\dis{C_A}} \Bigg\{ - \frac{3}{16} {\zeta_2}^2 \Bigg\} \,, 
                                     \nonumber\\
  {\overline {\cal G}}^{H,1}_{g,2} &=  {\dis{C_A n_f}}  \Bigg\{ - \frac{328}{81} + \frac{70}{9} \zeta_2 + \frac{32}{3} \zeta_3 \Bigg\}
                                     +  {\dis{C_A^{2}}} \Bigg\{ \frac{2428}{81} - \frac{469}{9} \zeta_2 
                                     + 4 {\zeta_2}^2 - \frac{176}{3} \zeta_3 \Bigg\} \,, 
                                     \nonumber\\
  {\overline {\cal G}}^{H,2}_{g,2} &=   {\dis{C_A^{2}}} \Bigg\{
                                     \frac{11}{40} {\zeta_2}^2  -
                                     \frac{203}{3} {\zeta_2} {\zeta_3}
                                     + \frac{1414}{27} {\zeta_2} +
                                     \frac{2077}{27} {\zeta_3}  + 43
                                     {\zeta_5}  - \frac{7288}{243}  \Bigg\} 
                                     \nonumber\\
                                   & +  {\dis{C_A n_f}} \Bigg\{
                                     -\frac{1}{20} {\zeta_2}^2 -
                                     \frac{196}{27} {\zeta_2} -
                                     \frac{310}{27} {\zeta_3} +  \frac{976}{243} \Bigg\}\, ,
                                     \nonumber\\
  {\overline {\cal G}}^{H,1}_{g,3} &= 
                                     {\dis{C_A}^3} \Bigg\{\frac{152}{63}
                                     \;{\zeta_2}^3  + \frac{1964}{9} \;{\zeta_2}^2
                                     + \frac{11000}{9} \;{\zeta_2}
                                     {\zeta_3}  - \frac{765127}{486} \;{\zeta_2}
                                     +\frac{536}{3} \;{\zeta_3}^2 -  \frac{59648}{27} \;{\zeta_3} 
                                     \nonumber\\
                                   &- \frac{1430}{3} \;{\zeta_5}
                                     +\frac{7135981}{8748}\Bigg\}
                                     + {\dis{C_A}^{2} {n_f}} 
                                     \Bigg\{-\frac{532}{9}
                                     \;{\zeta_2}^2 -   \frac{1208}{9} \;{\zeta_2} {\zeta_3}
                                     +\frac{105059}{243} \;{\zeta_2} +  \frac{45956}{81} \;{\zeta_3} 
                                     \nonumber\\
                                   &+\frac{148}{3} \;{\zeta_5} - \frac{716509}{4374} \Bigg\}
                                     +  {\dis{C_{A} {C_F} {n_f}}} \
                                     \Bigg\{\frac{152}{15} \;{\zeta_2}^2 
                                     - 88 \;{\zeta_2} {\zeta_3} 
                                     +\frac{605}{6} \;{\zeta_2} + \frac{2536}{27} \;{\zeta_3}
                                     +\frac{112}{3} \;{\zeta_5} 
                                     \nonumber\\
                                   &- \frac{42727}{324}\Bigg\}
                                     +  {\dis{C_{A} {n_f}^2}} \
                                     \Bigg\{\frac{32}{9} \;{\zeta_2}^2 - \frac{1996}{81} \;{\zeta_2}
                                     -\frac{2720}{81} \;{\zeta_3} + \frac{11584}{2187}\Bigg\}   \,.
\end{align}
The above ${\g}^{H,k}_{g.i}$'s enable us to compute the $\Phi^{A}_{g}$
up to three loop level. This completes all the ingredients required to
compute the SV cross section up to N$^{3}$LO that are presented in the
next section.

\section{SV Cross Sections}
\label{sec:Res}

In this section, we present our findings of the SV cross section at
N$^{3}$LO along with the results of previous orders. Expanding the SV
cross section $\Delta^{A, {\rm SV}}_{g}$, Eq.~(\ref{eq:sigma}), in
powers of $a_{s}$, we obtain
\begin{align}
  \label{eq:SVRenExp}
  \Delta_{g}^{A, {\rm SV}}(z, q^{2}, \mu_{R}^{2}, \mu_{F}^{2}) =
  \sum_{i=0}^\infty a_s^i \Delta_{g,i}^{A, {\rm SV}} (z, q^{2}, \mu_R^2, \mu_{F}^{2}) 
\end{align}
where,
\begin{align}
  \nn
  \Delta_{g,i}^{A, {\rm SV}} =
  \Delta_{g,i}^{A, {\rm SV}}|_\delta
  \delta(1-z) 
  + \sum_{j=0}^{2i-1} 
  \Delta_{g,i}^{A, {\rm SV}}|_{{\cal D}_j}
  {\cal D}_j \, .
\end{align}
Here, we present the results of the pseudo-scalar production cross
section up to
N$^{3}$LO for the choices of the scale
$\mu_{R}^{2}=\mu_{F}^{2}=q^{2}$ for which the Eq.~(\ref{eq:SVRenExp})
reads
\begin{align}
  \label{eq:SVRenExpAlt}
  \Delta_{g}^{A, {\rm SV}}(z, q^{2}) =
  \sum_{i=0}^\infty a_s^i(q^{2}) \Delta_{g,i}^{A, {\rm SV}} (z, q^{2})\,. 
\end{align}
with the following $\Delta_{g,i}^{A, {\rm SV}} (z, q^{2})$:
\begin{align}
  \label{eq:CIRes}
  \Delta^{A, {\rm SV}}_{g,0} &= {\dis{\delta(1-z)}}\,,
\nonumber\\
  \Delta^{A, {\rm SV}}_{g,1} &=  {\dis{\delta(1-z)}} \Bigg[
                               {\dis{C_{A}}} \Bigg\{ 8  + 8 \zeta_2 \Bigg\}
                               \Bigg]
                               +
                               {\dis{{\cal D}_{1}}} \Bigg[ {\dis{C_{A}}} \Bigg\{ 16\Bigg\}
                               \Bigg]\,,
                               \nonumber\\ 
  \Delta^{A, {\rm SV}}_{g,2} &= {\dis{\delta(1-z)}} \Bigg[{\dis{C_{A}^2}} \Bigg\{ \frac{494}{3} +
                               \frac{1112}{9} \zeta_2  - \frac{4}{5}
                               \zeta_2^2  - \frac{220}{3} \zeta_3
                               \Bigg\}
                               + 
                               {\dis{C_{A} n_{f}}} \Bigg\{ - \frac{82}{3}   -
                               \frac{80}{9} \zeta_2  - \frac{8}{3}
                               \zeta_3 \Bigg\}
                               \nonumber\\
                             &+ 
                               {\dis{C_{F} n_{f}}} \Bigg\{ - \frac{160}{3} + 12
                               \ln \left(\frac{q^2}{m_t^2}\right)
                               + 16 \zeta_3 \Bigg\}\Bigg]
                               + 
                               {\dis{{\cal D}_{0}}} \Bigg[ {\dis{C_{A} n_{f}}}
                               \Bigg\{ \frac{224}{27}  - \frac{32}{3}
                               \zeta_2 \Bigg\}  
                               \nonumber\\
                             &+ 
                               {\dis{C_{A}^2}} \Bigg\{ -
                               \frac{1616}{27}  + \frac{176}{3}
                               \zeta_2  + 312 \zeta_3 \Bigg\}
                               \Bigg]
                               + 
                               {\dis{{\cal D}_{1}}} \Bigg[ {\dis{C_{A} n_{f}}}
                               \Bigg\{ 
                               -
                               \frac{160}{9}  \Bigg\}
                               +
                               {\dis{C_{A}^2}} \Bigg\{ \frac{2224}{9}  - 160 \zeta_2 \Bigg\}
                               \Bigg] 
                               \nonumber\\
                             &+ {\dis{{\cal D}_{2}}} \Bigg[ {\dis{C_{A}^2}} \Bigg\{ -
                               \frac{176}{3} \Bigg\} 
                               +  
                               {\dis{C_{A} n_{f}}} \Bigg\{ \frac{32}{3} \Bigg\}
                               \Bigg]
                               + {\dis{{\cal D}_{3}}} \Bigg[ C_{A}^2 \Bigg\{ 128 \Bigg\}
                               \Bigg]\,,
                               \nonumber\\ 
  \Delta^{A, {\rm SV}}_{g,3} &= {\dis{\delta(1-z)}} \Bigg[ {\dis{n_{f}
                               C_{J}^{(2)}}} \Bigg\{ -4 \Bigg\}
                               + 
                               {\dis{C_{F} n_{f}^2}} \Bigg\{
                               \frac{1498}{9}  - \frac{40}{9} \zeta_2
                               - \frac{32}{45} \zeta_2^2  - 
                               \frac{224}{3} \zeta_3 \Bigg\} 
                               \nonumber\\
                             &+ {\dis{C_{A}^3}}
                               \Bigg\{ \frac{114568}{27}  +
                               \frac{137756}{81} \zeta_2  - 
                               \frac{61892}{135} \zeta_2^2 -
                               \frac{64096}{105} \zeta_2^3  - 3932 \zeta_3 + 
                               \frac{7832}{3} \zeta_2 \zeta_3 
                               \nonumber\\
                             &+
                               \frac{13216}{3} \zeta_3^2  -
                               \frac{30316}{9} \zeta_5 \Bigg\}  
                               + 
                               {\dis{C_{F}^2 n_{f}}} \Bigg\{ \frac{457}{3} +
                               208 \zeta_3  - 320 \zeta_5 \Bigg\} 
                               + 
                               {\dis{C_{A}^2 n_{f}}} \Bigg\{ -
                               \frac{113366}{81} 
                               \nonumber\\
                             &- \frac{10888}{81}
                               \zeta_2  + \frac{21032}{135} \zeta_2^2 + 
                               \frac{8840}{27} \zeta_3 -
                               \frac{2000}{3} \zeta_2 \zeta_3  +
                               \frac{6952}{9} \zeta_5 \Bigg\}  
                               + 
                               {\dis{C_{A} n_{f}^2}} \Bigg\{
                               \frac{6914}{81} - \frac{1696}{81}
                               \zeta_2  
                               \nonumber\\
                             &- \frac{608}{45} \zeta_2^2 +
                               \frac{688}{27} \zeta_3 \Bigg\}  
                               + 
                               {\dis{C_{A} C_{F} n_{f}}} \Bigg\{ - 1797 -
                               \frac{4160}{9} \zeta_2  + 
                               \frac{176}{45} \zeta_2^2 +
                               \frac{1856}{3} \zeta_3 + 192 \zeta_2
                               \zeta_3  
                               \nonumber\\
                             &+ 160 \zeta_5 + 96 \ln
                               \left(\frac{q^2}{m_t^2}\right)  +
                               96  \ln \left(\frac{q^2}{m_t^2}\right) \zeta_2 \Bigg\} 
                               \Bigg]
                               +
                               {\dis{{\cal D}_{0}}} \Bigg[ 
                               {\dis{C_{A}^2 n_{f}}}
                               \Bigg\{ \frac{173636}{729}  -
                               \frac{41680}{81} \zeta_2  
                               \nonumber\\
                             &-
                               \frac{544}{15} \zeta_2^2  -
                               \frac{7600}{9} \zeta_3 \Bigg\}  
                               + 
                               {\dis{C_{A} C_{F} n_{f}}} \Bigg\{
                               \frac{3422}{27} - 32 \zeta_2  -
                               \frac{64}{5} \zeta_2^2  -
                               \frac{608}{9} \zeta_3 \Bigg\}  
                               \nonumber\\
                             &+ 
                               {\dis{C_{A} n_{f}^2}} \Bigg\{ - \frac{3712}{729} +
                               \frac{640}{27} \zeta_2  +
                               \frac{320}{27} \zeta_3 \Bigg\} 
                               + 
                               {\dis{C_{A}^3}} \Bigg\{ - \frac{943114}{729} +
                               \frac{175024}{81} \zeta_2  +
                               \frac{4048}{15} \zeta_2^2  
                               \nonumber\\
                             &+ 
                               \frac{210448}{27} \zeta_3 -
                               \frac{23200}{3} \zeta_2 \zeta_3   + 11904 \zeta_5 \Bigg\}
                               \Bigg]
                               +
                               {\dis{{\cal D}_{1}}} \Bigg[ 
                               {\dis{C_{A}^3}} \Bigg\{
                               \frac{414616}{81} - \frac{13568}{3}
                               \zeta_2  - \frac{9856}{5} \zeta_2^2 
                               \nonumber\\
                             &-
                               \frac{22528}{3} \zeta_3 \Bigg\}  
                               + 
                               {\dis{C_{A}^2 n_{f}}} \Bigg\{ -
                               \frac{79760}{81} + \frac{6016}{9}
                               \zeta_2  + \frac{2944}{3} \zeta_3
                               \Bigg\} 
                               + 
                               {\dis{C_{A} n_{f}^2}} \Bigg\{
                               \frac{1600}{81} - \frac{256}{9}
                               \zeta_2 \Bigg\}  
                               \nonumber\\
                             &+ {\dis{C_{A} C_{F} n_{f}}} \Bigg\{ -
                               1000  + 384 \zeta_3 + 192 \ln \left(\frac{q^2}{m_t^2}\right) \Bigg\}
                               \Bigg] 
                               +
                               {\dis{{\cal D}_{2}}} \Bigg[
                               {\dis{C_{A} C_{F} n_{f}}}
                               \Bigg\{ 32 \Bigg\} 
                               \nonumber\\
                             &+ {\dis{C_{A} n_{f}^{2}}}
                               \Bigg\{ - \frac{640}{27} \Bigg\}  
                               +
                               {\dis{C_{A}^2 n_{f}}} \Bigg\{ \frac{16928}{27}
                               - \frac{2176}{3} \zeta_2 \Bigg\} 
                               + 
                               {\dis{C_{A}^3}} \Bigg\{ - \frac{79936}{27} +
                               \frac{11968}{3} \zeta_2  
                               \nonumber\\
                             &+ 11584 \zeta_3 \Bigg\}
                               \Bigg]
                               + 
                               {\dis{{\cal D}_{3}}} \Bigg[ 
                               {\dis{C_{A}^2 n_{f}}} \Bigg\{ -
                               \frac{10496}{27} \Bigg\} 
                               + {\dis{C_{A} n_{f}^2}} \Bigg\{ \frac{256}{27} \Bigg\}
                               + {\dis{C_{A}^3}} \Bigg\{
                               \frac{86848}{27}  
                               \nonumber\\
                             &- 3584 \zeta_2 \Bigg\}
                               \Bigg]
                               + {\dis{{\cal D}_{4}}} \Bigg[ 
                               {\dis{C_{A}^3}} \Bigg\{ - \frac{7040}{9} \Bigg\} + {\dis{C_{A}^2
                               n_{f}}} \Bigg\{ \frac{1280}{9}\Bigg\}
                               \Bigg] 
                               +
                               {\dis{{\cal D}_{5}}} \Bigg[ {\dis{C_{A}^3}} \Bigg\{ 512 \Bigg\}
                               \Bigg]\,.
\end{align}
The SV cross section up to NNLO are in agreement with the existing
ones, computed in the
article~\cite{Ravindran:2003um, Harlander:2002vv, Anastasiou:2002wq}.

\section{Threshold Resummation}
\label{sec:Resum}

Despite the spectacular accuracy of the fixed order results which are
defined in power series expansions of the strong coupling constant
$a_{s}$, it is necessary, in certain cases, to resum the dominant
contributions to all orders in $a_{s}$ to get more reliable
predictions and to reduce the scale uncertainties significantly. In
case of threshold corrections, due to soft-gluon emission the fixed
order pQCD calculation may yield large threshold logarithms of the
kind ${\cal D}_{i}$, defined in Eq.~(\ref{eq:calD}), hence we must
resum these contributions to all orders in $a_{s}$. The resummation of
these so-called Sudakov logarithms is usually pursued in Mellin space
using the formalism developed in \cite{Sterman:1986aj, Catani:1989ne,
  Catani:1996yz, Contopanagos:1996nh}. Alternatively, it is performed
in the framework of soft-collinear effective field theory
(SCET)~\cite{hep-ph/0005275, hep-ph/0011336, hep-ph/0107001,
  hep-ph/0109045, hep-ph/0206152, hep-ph/0211358,
  hep-ph/0202088}. Here, we will discuss this in the context of Mellin
space formalism. 

\subsection{Mellin Space Prescription}
\label{ss:MellinSpRes}

Under this prescription, the threshold resummation is performed in
Mellin-$N$ space where the $N$-th order Mellin moment is defined with
respect to the partonic scaling variable $z$. In Mellin space, the
threshold limit $z\rightarrow 1$ corresponds to $N\rightarrow \infty$
and the plus distributions ${\cal D}_{i}$, Eq.~(\ref{eq:calD}), take
the form $\ln^{i-1} N$.  These logarithmic contributions are evaluated
to all orders in perturbation theory by performing the threshold
resummation through~\cite{Sterman:1986aj, Catani:1989ne,
  Catani:1996yz, Contopanagos:1996nh}
\begin{align}
  \label{eq:DeltaResum}
  \Delta^{A,{\rm res}}_{g,N}(q^{2}, \mu_{R}^{2}, \mu_{F}^{2}) &= C^{A, {\rm th}}_{g}(q^{2},
                                                                \mu_{R}^{2},
                                                                \mu_{F}^{2})
                                                                \Delta_{g,N}(q^{2})\,.
\end{align}
The component $C^{A, {\rm th}}_{g}$ depends on both the initial as
well as final state particles, though it is independent of the
variable $N$. On the other hand, the remaining part $\Delta_{g,N}$
does not care about the details of the final state particle, it only
depends on the initial state partons and the variable $N$. Being
independent of the nature of the final state, $\Delta_{g,N}$ can be
considered as a universal quantity which is same for any operator. In
addition, it is investigated in the articles \cite{Sterman:1986aj,
  Catani:1989ne} that it arises solely from the soft parton radiation
and it resums all the perturbative contributions $a_{s}^{n} \ln^{m}N$
($m \geq 0$) to all orders.  Our goal is to calculate the threshold
resummation factor $C^{A,{\rm th}}_{g}$ which encapsulates all the
remaining $N$-independent contributions to the resummed partonic cross
section~\ref{eq:DeltaResum}. Below, we demonstrate the prescription
based on our formalism to calculate this quantity $C^{A,{\rm th}}_{g}$
order by order in perturbation theory.

In the article~\cite{Ravindran:2006cg}, it was shown how the
soft-collinear distribution $\Phi^{A}_{g}(=\Phi_{g})$,
Eq.~(\ref{eq:PhiSoln}), captures all the features of the $N$-space
resummation. In this section, we discuss that prescription briefly in
the present context. Using the well known identity
\begin{align}
  \label{eq:IdDeltaD}
  \frac{1}{1-z} \left[ (1-z)^{2} \right]^{j \frac{\epsilon}{2}} =
  \frac{1}{j\epsilon} \delta(1-z) + \left( \frac{1}{1-z} \left[
  (1-z)^{2} \right]^{j \frac{\epsilon}{2}} \right)_{+}\,,
\end{align}
we can express the soft-collinear distribution~\ref{eq:PhiSoln} as
\begin{align}
  \label{eq:PhiResum}
  \Phi^{A}_{g} &= \left( \frac{1}{1-z} \Bigg\{
                 \int_{\mu_{R}^{2}}^{q^{2}(1-z)^{2}}
                 \frac{d\lambda^{2}}{\lambda^{2}}
                 A^{A}_{g} \l(a_{s}(\lambda^{2}) \r)  
                 + {\overline G}^{A}_{g}\left(
                 a_{s}(q^{2}(1-z)^{2}),\epsilon \right) \Bigg\}
                 \right)_{+} 
                 \nonumber\\
               &+ \delta(1-z) \sum_{j=1}^{\infty} {\hat a}_{s}^{j} \l(
                 \frac{q^2}{\mu^2} \r)^{j \frac{\epsilon}{2}}
                 S^{j}_{\epsilon} {\hat \phi}^{A}_{g,j} (\epsilon)
                 + \left( \frac{1}{1-z} \right)_{+} \sum_{j=1}^{\infty} {\hat
                 a}_{s}^{j}  \l(
                 \frac{\mu_{R}^2}{\mu^2} \r)^{j \frac{\epsilon}{2}}
                 S^{j}_{\epsilon} {\overline K}^{A}_{g,j}(\epsilon)
\end{align}
where, all the quantities are already introduced in
Sec.~\ref{sec:ThreResu} except ${\overline K}^{A}_{g,j}(\epsilon)$
which is defined through the expansion of ${\overline K}^{A}_{g}$,
appeared in Eq.~(\ref{eq:KGbarEqn}), in powers of ${\hat a}_{s}$ in
the following way:
\begin{align}
  \label{eq:Kbarj}
  {\overline K}^{A}_{g}\left( {\hat a}_{s}, \frac{\mu_{R}^{2}}{\mu^{2}},
  z, \epsilon \right) &= \delta(1-z) \sum_{j=1}^{\infty} {\hat
                        a}_{s}^{j} \left( \frac{\mu_{R}^{2}}{\mu^{2}}
                        \right)^{j \frac{\epsilon}{2}}
                        S^{j}_{\epsilon} {\overline K}^{A}_{g,j}(\epsilon)\,.
\end{align}
The identification of the first plus distribution part of
$\Phi^{A}_{g}$, Eq.~(\ref{eq:PhiResum}), with the factor contributing
to the process independent $\Delta_{g,N}(q^{2})$ has been discussed in
the same article~\cite{Ravindran:2006cg} which reads
\begin{align}
  \label{eq:ResumDeltagN}
  \Delta_{g,N} &= \exp \left[ \int_{0}^{1} dz
                 \frac{z^{N-1}-1}{1-z} \Bigg\{ 2
                 \int_{q^{2}}^{q^{2}(1-z)^{2}}
                 \frac{d\lambda^{2}}{\lambda^{2}}
                 A_{g} \l(a_{s}(\lambda^{2}) \r)  
                 + D_{g}\left(
                 a_{s}(q^{2}(1-z)^{2})\right) \Bigg\} \right] 
\end{align}
with
\begin{align}
  \label{eq:RelnGD}
  D_{g}\left( a_{s}(q^{2}(1-z)^{2}) \right) &= 2 {\overline G}_{g}\left(
                                              a_{s}(q^{2}(1-z)^{2}),\epsilon \right)|_{\epsilon=0}\,.
\end{align}
In the above expression, the superscript $A$ has been omitted to
emphasise the universal nature of these quantities.  The remaining
part of the Eq.~(\ref{eq:PhiResum}) along with the other parts,
namely, form factor, operator renormalisation constant and mass
factorisation kernel in Eq.~(\ref{eq:psi}) contribute to
$C^{A,{\rm th}}_{g}$. Expanding this in powers of $a_{s}$ as
\begin{align}
  \label{eq:CAthExpand}
  C^{A,{\rm th}}_{g} &= 1+\sum_{j=1}^{\infty} a_{s}^{j}\, C^{A,{\rm th}}_{g,j}\,,
\end{align}
we determine $C^{A,{\rm th}}_{g,j}$ up to three loop ($j=3$) order
which are provided below (with the choice
$\mu_{R}^{2}=\mu_{F}^{2}=q^{2}$):

\begin{align}
  \label{eq:Cth}
  C^{A,{\rm th}}_{g,1} &= {\dis{C_{A}}} \Bigg\{ 8 + 8 \zeta_2 \Bigg\}\,,
                         \nonumber\\ 
  C^{A,{\rm th}}_{g,2} &= {\dis{C_{A}^2}} \Bigg\{ \frac{494}{3} +
                         \frac{1112}{9} \zeta_2  + 12 \zeta_2^2 -
                         \frac{220}{3} \zeta_3 \Bigg\} 
                         + 
                         {\dis{C_{A} n_{f}}} \Bigg\{ - \frac{82}{3} - \frac{80}{9} \zeta_2 -
                         \frac{8}{3} \zeta_3 \Bigg\} 
                         \nonumber\\
                       &+ 
                         {\dis{C_{F} n_{f}}} \Bigg\{ - \frac{160}{3} + 12 \ln
                         \left(\frac{q^2}{m_t^2}\right)  + 16 \zeta_3 \Bigg\}\,,
                         \nonumber\\ 
  C^{A,{\rm th}}_{g,3} &=   {\dis{n_{f} C^{(2)}_{J}}} \Bigg\{ -4 \Bigg\} 
                         + 
                         {\dis{C_{F} n_{f}^2}} \Bigg\{ \frac{1498}{9} -
                         \frac{40}{9} \zeta_2  - \frac{32}{45} \zeta_2^2 - 
                         \frac{224}{3} \zeta_3 \Bigg\} 
                         + 
                         {\dis{C_{F}^2 n_{f}}} \Bigg\{ \frac{457}{3} + 208
                         \zeta_3  
                         \nonumber\\
                       &- 320 \zeta_5 \Bigg\} 
                         + 
                         {\dis{C_{A}^2 n_{f}}} \Bigg\{ - \frac{113366}{81} -
                         \frac{10888}{81} \zeta_2  + \frac{17192}{135}
                         \zeta_2^2  + 
                         \frac{584}{3} \zeta_3 - \frac{464}{3} \zeta_2
                         \zeta_3  +
                         \frac{808}{9} \zeta_5 \Bigg\} 
                         \nonumber\\
                       &+ 
                         {\dis{C_{A}^3}} \Bigg\{ \frac{114568}{27} +
                         \frac{137756}{81} \zeta_2  - \frac{4468}{27}
                         \zeta_2^2  - \frac{32}{5} \zeta_2^3 - 
                         \frac{80308}{27} \zeta_3 - \frac{616}{3}
                         \zeta_2 \zeta_3  + 96 \zeta_3^2 
                         \nonumber\\
                       &+ 
                         \frac{3476}{9} \zeta_5 \Bigg\} 
                         + 
                         {\dis{C_{A}  n_{f}^2}} \Bigg\{ \frac{6914}{81}  - \frac{1696}{81} \zeta_2
                         - \frac{608}{45} \zeta_2^2 + \frac{688}{27}
                         \zeta_3 \Bigg\} 
                         + 
                         {\dis{C_{A} C_{F} n_{f}}}  \Bigg\{ - 1797   
                         \nonumber\\
                       &+
                         96 \ln  \left(\frac{q^2}{m_t^2}\right) -
                         \frac{4160}{9} \zeta_2  + 96 \ln
                         \left(\frac{q^2}{m_t^2}\right) \zeta_2  + 
                         \frac{176}{45} \zeta_2^2 + \frac{1856}{3}
                         \zeta_3 + 192 \zeta_2 \zeta_3  
                         \nonumber\\
                       &+ 160 \zeta_5 \Bigg\}\,.
\end{align}
The above new result of $C^{A,{\rm th}}_{g,3}$ along with the
universal factor $\Delta_{g,N}$ provide the threshold resummed cross
section of the pseudo-scalar production at N$^{3}$LL accuracy. The
more elaborate discussion on this prescription to perform threshold
resummation will be presented elsewhere by us.

\section{Numerical Impact of SV Cross Section}
\label{sec:Disc}

In this section, we present our findings on the numerical impact of
threshold N$^3$LO predictions in QCD for the production of a
pseudo-scalar Higgs boson at the LHC and also make comparison with the
corresponding results for the SM Higgs boson.  As we are interested in
quantifying the QCD effects, we assume that pseudo-scalar couples only
to top quarks.  Hence, the dominant contribution resulting from bottom
quark initiated processes can be included in a systematic way in our
numerical study but we do not do it here.  Moreover, our predictions
are based on the effective theory approach where the top quarks are
integrated out and we have only light quarks.  Like in the case of
predictions for the Higgs production in the effective theory, for the
pseudo-scalar production we multiply the born cross section computed
using the finite top mass ($m_t=172.5$ GeV) with higher orders which
are obtained in the effective theory.  Without loss of generality, we
normalise the cross section by $\hbox{cot}^2\beta$.  The mass of the
pseudo-scalar is taken to be $m_A = 200$ GeV and the component of the
Wilson coefficient $C^{(2)}_J$ is considered to be zero due to its
unavailability in the literature. We use {\tt MSTW2008}~\cite{Martin:2009iq} parton
distribution functions (PDFs) throughout where the LO, NLO and NNLO
parton level cross sections are convoluted with the corresponding {\tt
  MSTW2208lo, MSTW2008nlo} and {\tt MSTW2008nnlo} PDFs while for
N$^3$LO$_{\rm SV}$ cross sections we use {\tt MSTW2008nnlo} PDFs.  The
strong coupling constant is provided by the respective PDFs from {\tt
  LHAPDF} with $\alpha_s(m_Z)$ = 0.1394(LO), 0.12018(NLO) and
0.11707(NNLO).

To estimate the impact of QCD corrections, we define the K-factors as
\begin{eqnarray}
\hbox{K}^{(1)} = \frac{ \sigma^{\rm NLO} } {\sigma^{\rm LO}},  \hspace{1.5cm}
\hbox{K}^{(2)} = \frac{ \sigma^{\rm NNLO} } {\sigma^{\rm LO}}, \hspace{1.5cm}
\hbox{K}^{(3)} = \frac{ \sigma^{\rm N^3LO_{\rm SV}} } {\sigma^{\rm LO}}
\label{eqn:kf}
\end{eqnarray}
\begin{figure*}[htb]
\centerline{
\epsfig{file=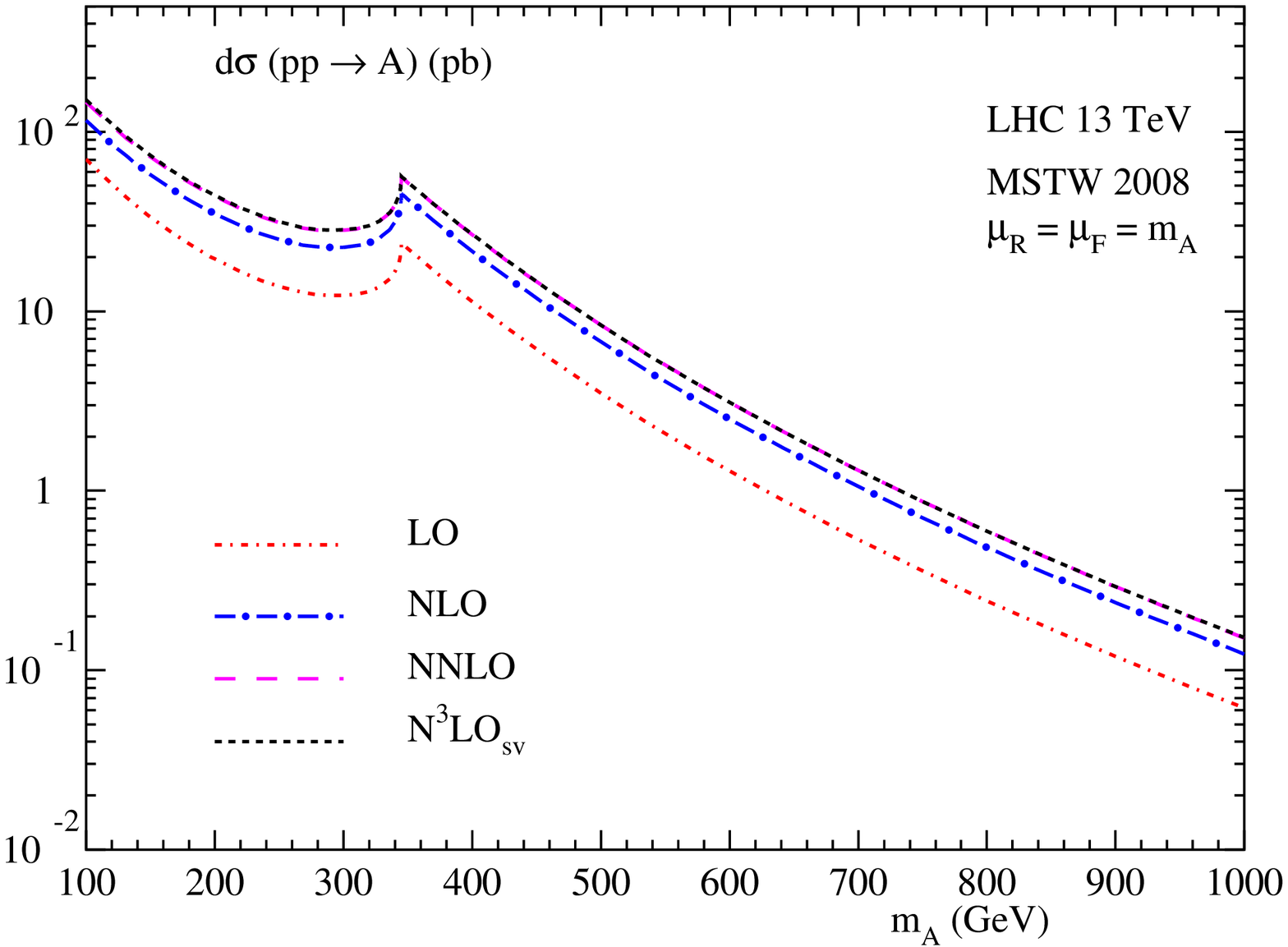,width=8cm,height=6.5cm,angle=0}
\epsfig{file=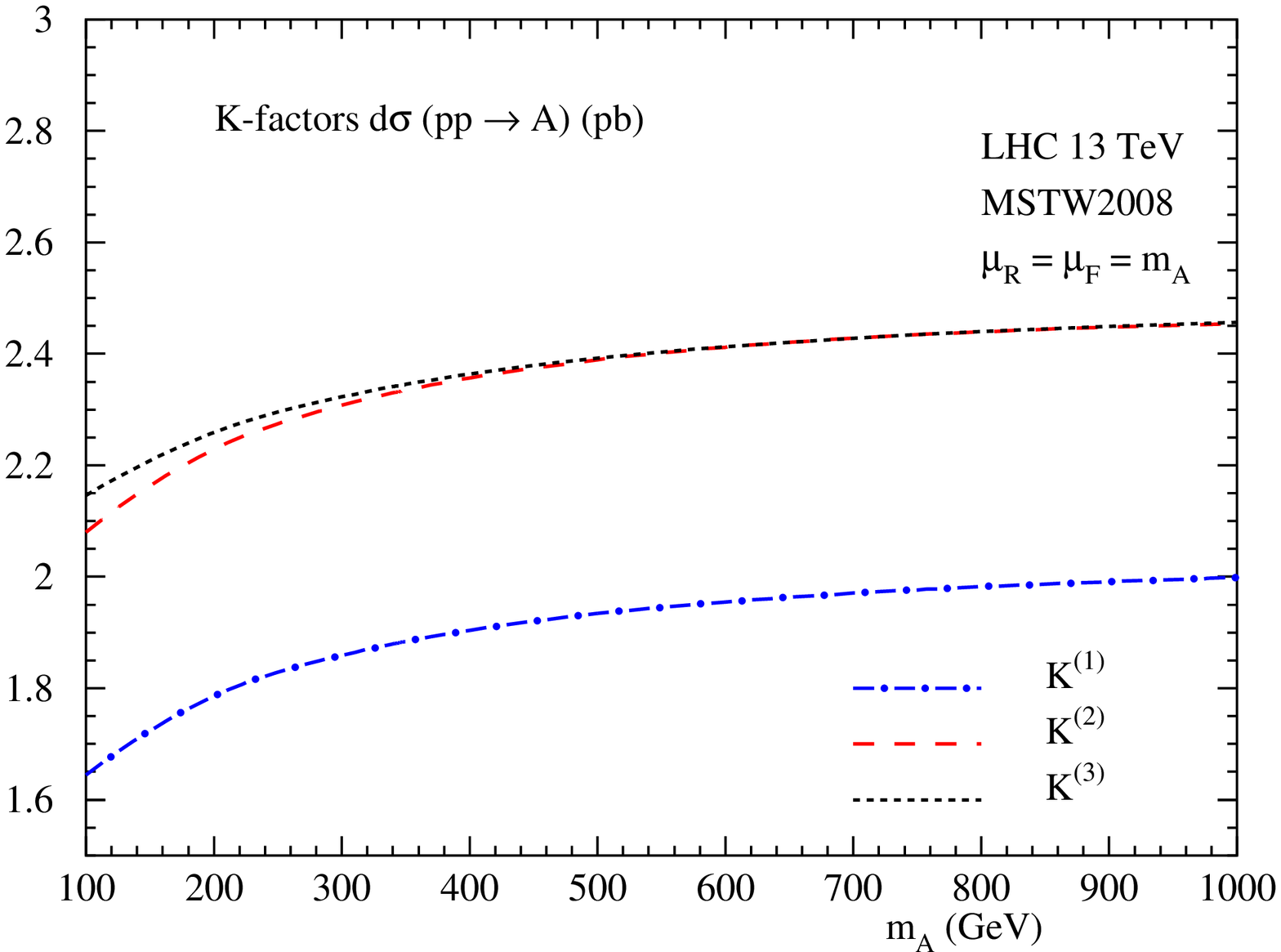,width=8cm,height=6.5cm,angle=0}
}
\caption{\sf Pseudo-scalar production cross section (left panel) for LHC13 
and the corresponding K-factors (right panel). The observed spike at $345$ 
GeV indicates the top quark pair threshold region.}
\label{mavar1}
\end{figure*}
\begin{figure*}[htb]
\centerline{
\epsfig{file=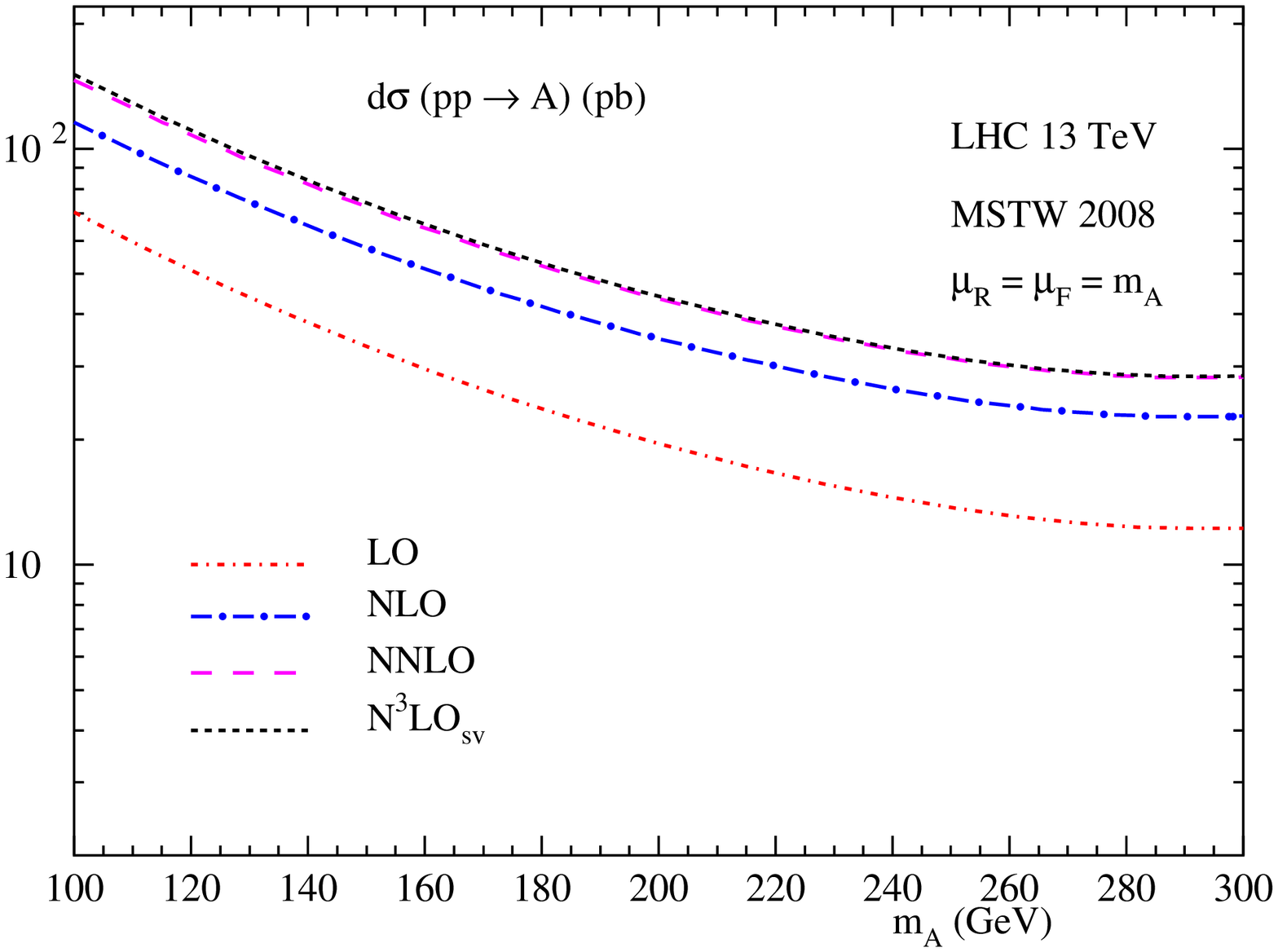,width=8cm,height=6.5cm,angle=0}
\epsfig{file=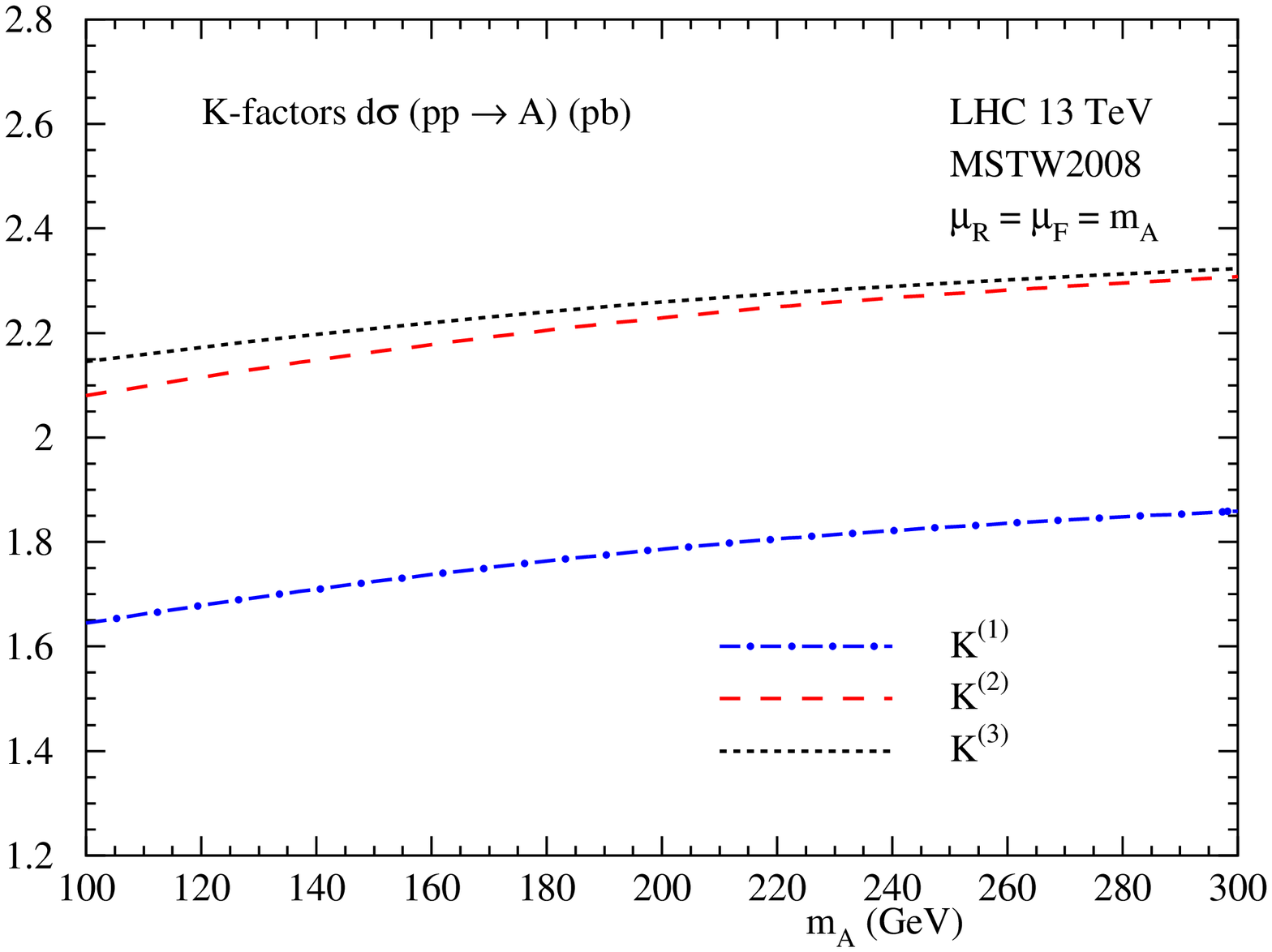,width=8cm,height=6.5cm,angle=0}
}
\caption{\sf Same as fig.~\ref{mavar1} but smaller values of $m_A$.}
\label{mavar2}
\end{figure*}
In fig.~\ref{mavar1}, for LHC13, we plot the pseudo-scalar production
cross section as a function of its mass $m_A$.  Since we retain the
dependence on the $m_t$ at the born level, beyond the top pair
threshold ($\tau_A > 1$), due to change in the functional dependence
of $\tau_A$ one finds a spike at $2m_t$ (left panel).  The
corresponding K-factors are given in the right panel and are in
general found to increase with $m_A$.  The NLO correction enhances the
LO predictions by as much as 100\% for $m_A=1$ TeV, whereas the NNLO
correction adds about an additional 45\%. On the other hand the
${\rm N^3LO}_{\rm SV}$ correction is found to be about 1.5\% of LO for
small mass region $m_A < 300$ GeV and for higher $m_A$ values the
correction at the ${\rm N^3LO}_{\rm SV}$ level becomes even smaller,
about 0.3\% for $m_A=1$ TeV.  In either case, these N$^3$LO$_{\rm SV}$
effects show a convergence of the perturbation series.

\begin{figure*}[h]
\centerline{
\epsfig{file=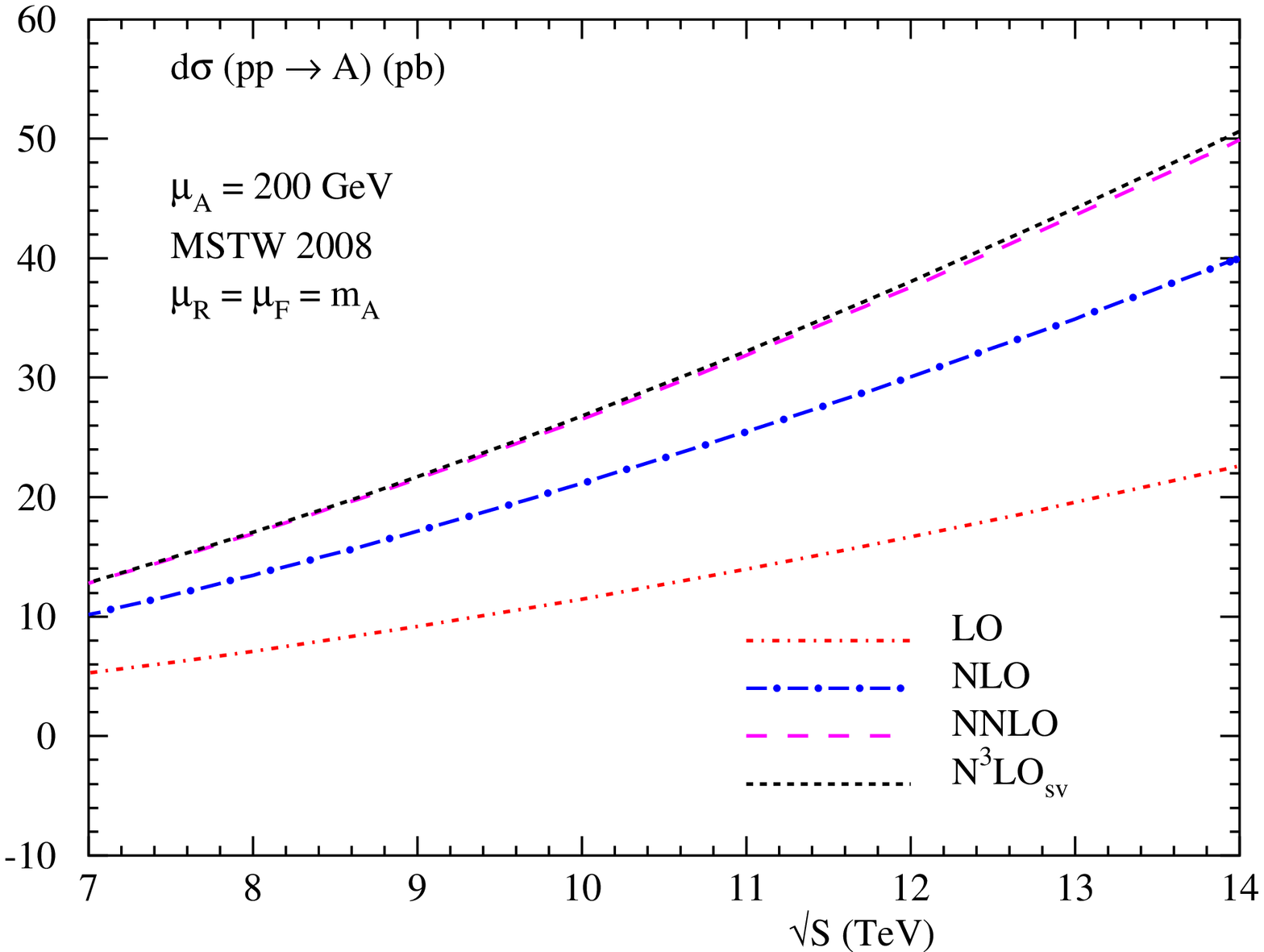,width=8cm,height=6.5cm,angle=0}
\epsfig{file=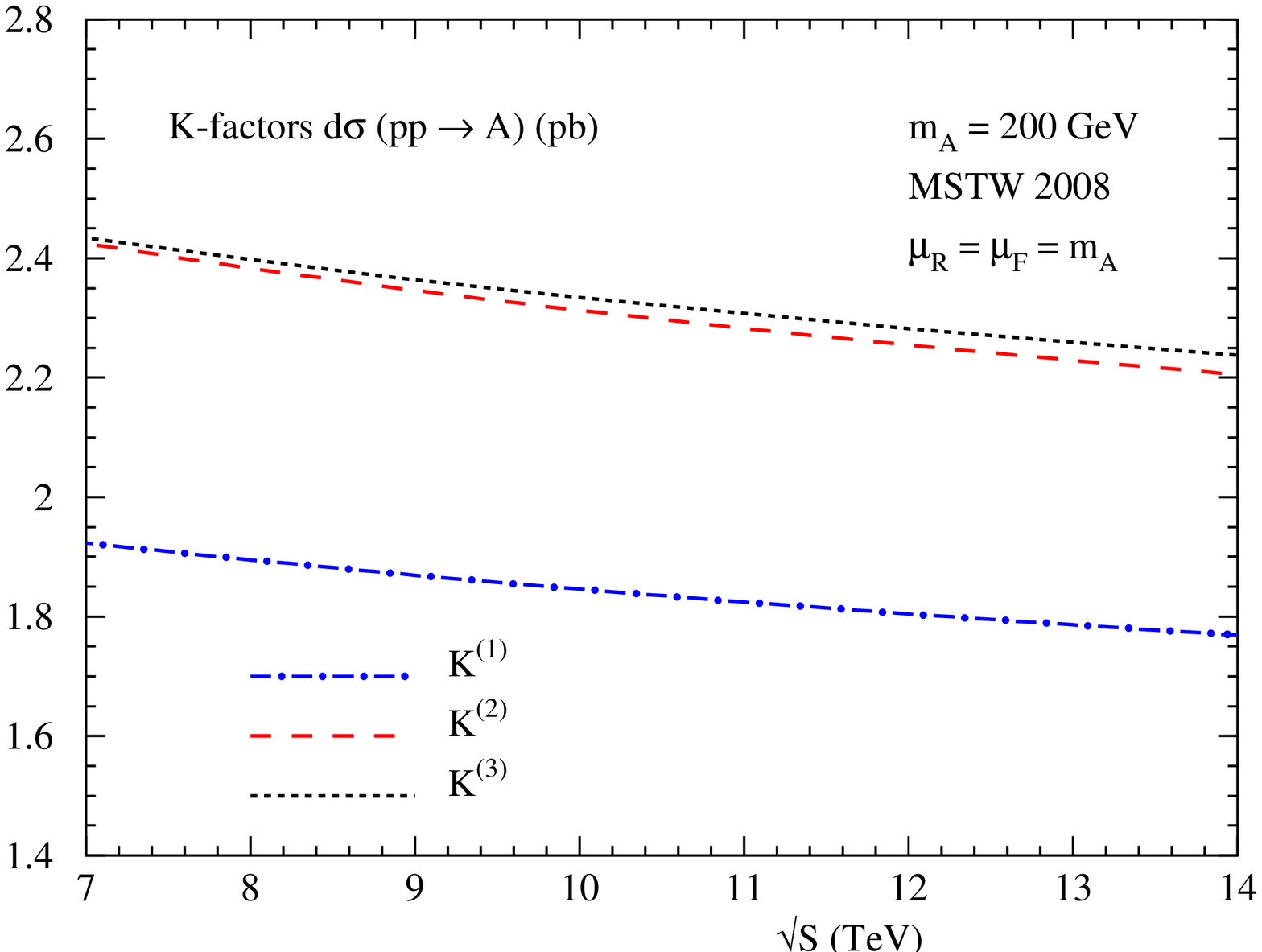,width=8cm,height=6.5cm,angle=0}
}
\caption{\sf Pseudo-scalar production cross section as a function of $\sqrt{S}$ 
(left panel) and the corresponding K-factors (right panel).}
\label{ecmvar}
\end{figure*}
In fig.~\ref{ecmvar}, we present the cross sections as a function of
the center of mass energy $\sqrt{S}$ of the incoming protons at the
LHC. The increase in the cross sections (left panel) with $\sqrt{S}$
is simply because of the increase in the corresponding parton fluxes
for any given $m_A$. On the contrary, the corresponding K-factors
(right panel) increase with decreasing $\sqrt{S}$ for fixed $m_A$.  A
similar pattern is shown both in figs.~(\ref{mavar1} \& \ref{mavar2})
where the K-factors increase with $m_A$ for a given $\sqrt{S}$.  The
guiding principle for the behaviour of the K-factors in these two
cases is the same, namely, as $m_A$ approaches $\sqrt{S}$, the cross
sections are dominated by large soft gluon effects.

\begin{figure*}[htb]
\centerline{
\epsfig{file=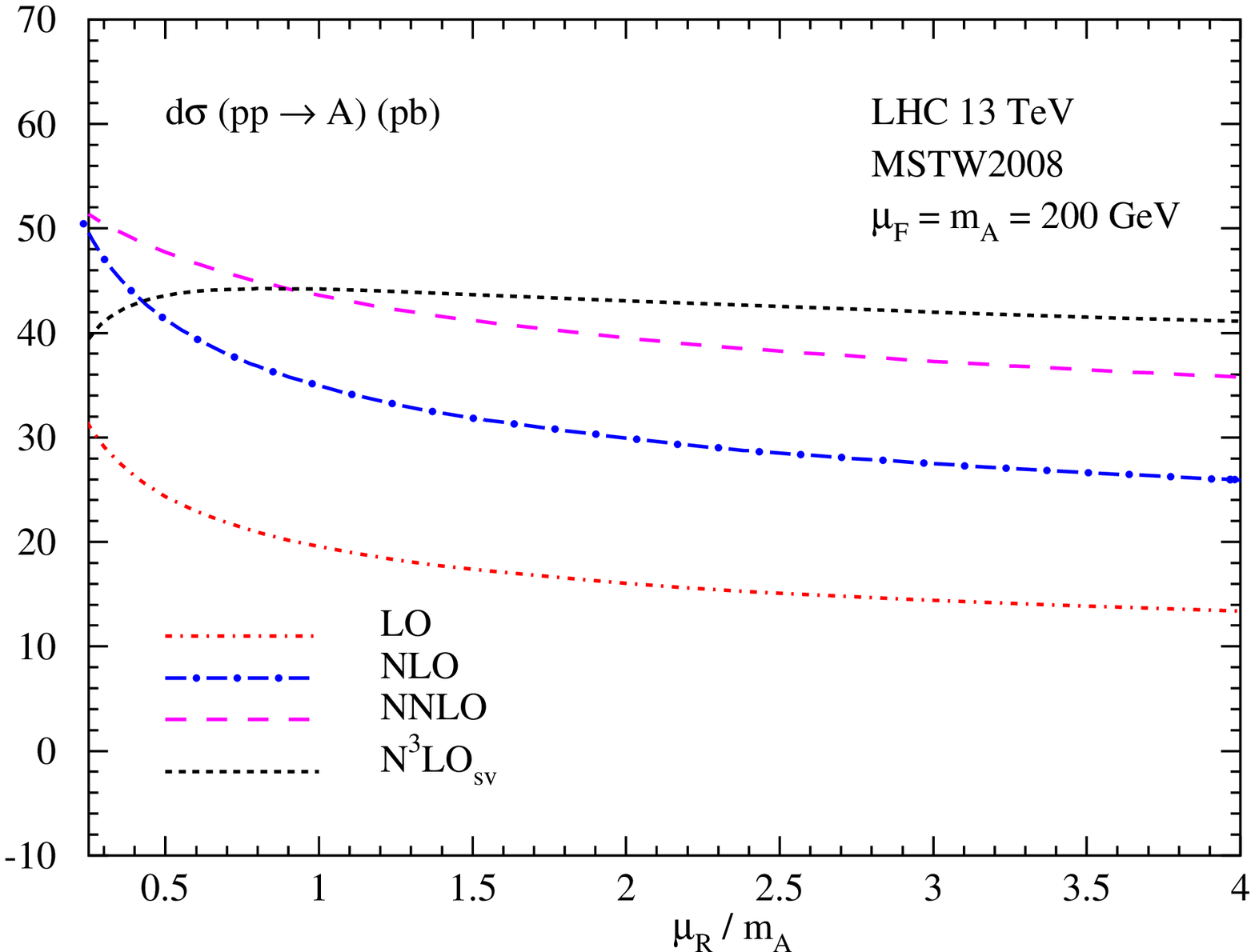,width=8cm,height=6.5cm,angle=0}
\epsfig{file=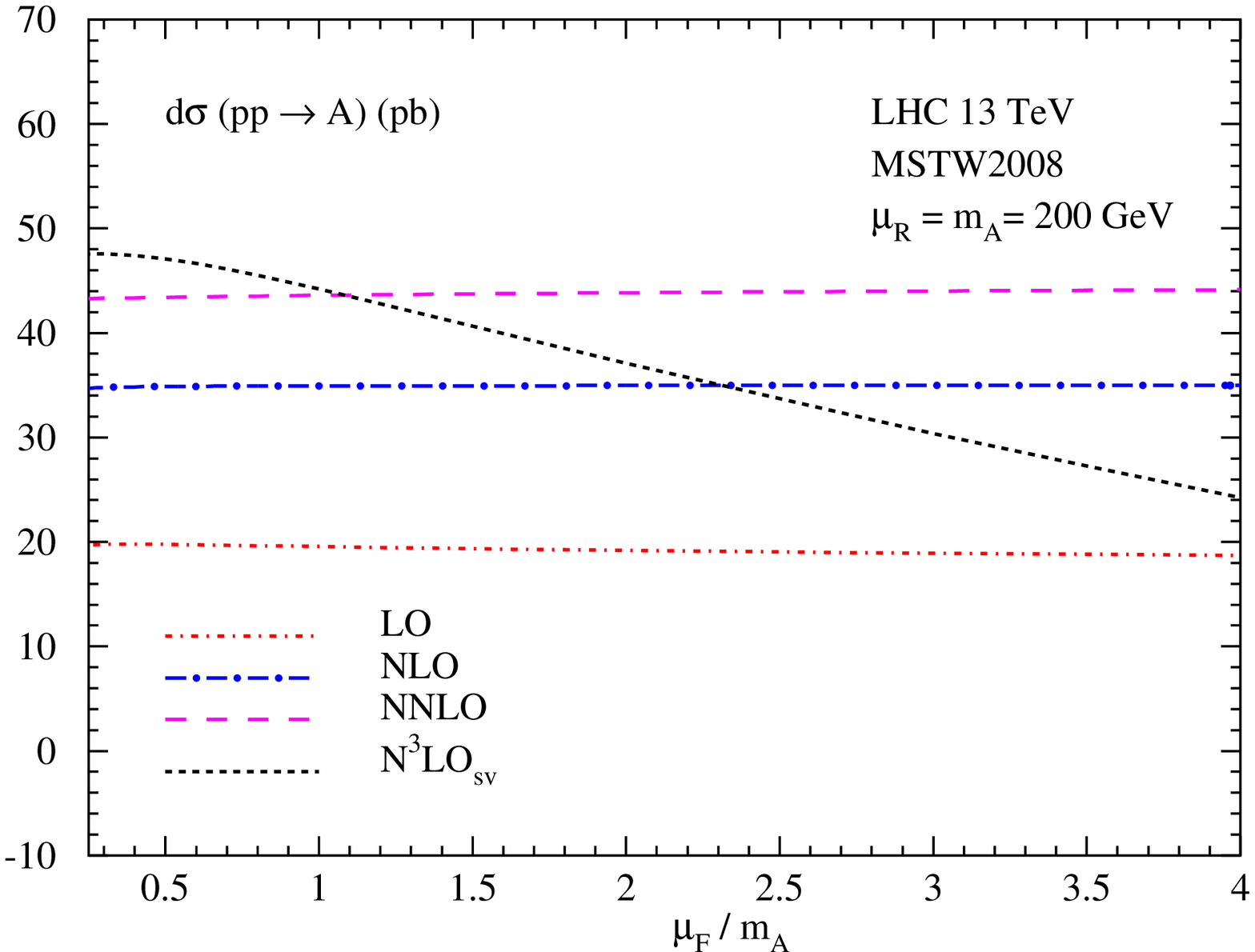,width=8cm,height=6.5cm,angle=0}
}
\caption{\sf Scale uncertainties associated with the pseudo-scalar production cross section 
for LHC13. Variation with $\mu_R$ keeping $\mu_F=m_A$ fixed (left panel). Variation with 
$\mu_F$ keeping $\mu_R = m_A$ fixed (right panel).}
\label{scalevarps}
\end{figure*}
\begin{figure*}[htb]
\centerline{
\epsfig{file=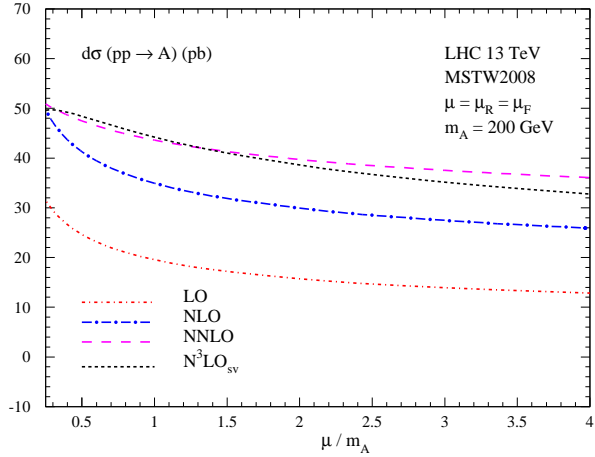,width=8cm,height=6.5cm,angle=0}
}
\caption{\sf Scale uncertainties associated with the pseudo-scalar production cross section 
for LHC13 with $\mu = \mu_R = \mu_F$.}
\label{murfps}
\end{figure*}
Next, we present the scale ($\mu_R,\mu_F$) uncertainties up to
${\rm N^3LO}_{\rm SV}$ in fig.~\ref{scalevarps} for the choice of
$m_A=200$ GeV. In the left panel, we vary the renormalisation scale
$\mu_R$ between $m_A/4$ and $4m_A$, keeping $\mu_F=m_A$ fixed. Unlike
the Drell-Yan process, for the pseudo-scalar production the
renormalisation scale $\mu_R$ enters even at LO through the strong
coupling constant $a_s$. This is identical to the SM Higgs boson
production in the gluon fusion channel.  This is the main source of
large scale uncertainty at LO. It gets significantly reduced when we
include NLO and NNLO corrections as expected and it continues to do so
at N$^3$LO level. In the right panel, we show the factorisation scale
uncertainties by varying $\mu_F$ from $m_A/4$ to $4m_A$ and fixing
$\mu_R=m_A$.  Here, the fixed order results show improvement in the
reduction of factorisation scale uncertainty from NLO to
NNLO. However, due to the lack of parton distribution functions at
N$^3$LO level and also due to the missing regular contributions from
the parton level cross sections, the SV corrections at three loop
level do not show any improvement of the factorisation scale
uncertainties.  In fig.~\ref{murfps}, we show the combined effect of
$\mu_R$ and $\mu_F$ scale uncertainties by varying the scale $\mu$
between $m_A/4$ and $4m_A$, where $\mu=\mu_R=\mu_F$.  Here, the NNLO
cross sections show a good improvement over the NLO ones, while the
scale uncertainties at N$^3$LO$_{\rm SV}$ are slightly larger but
comparable to the NNLO ones.

\begin{table}[h!]
\centering
\begin{tabular}{| c | c | c | c | c | c | c |}
    \hline
     \multirow{2}{*}{$\sqrt{S}$ TeV} & 
     \multicolumn{3}{c|}{SM Higgs} & 
     \multicolumn{3}{c|}{Pseudo-scalar} \\ 
     \cline{2-7}
     {} &  K$^{(1)}$ & K$^{(2)}$ & K$^{(3)}$ &K$^{(1)}$  & K$^{(2)}$  & K$^{(3)}$  \\
    \hline \hline
     7  & 1.83 & 2.31 & 2.44 & 1.84 & 2.34 & 2.37 \\
    \hline
     8  & 1.79 & 2.27 & 2.40 & 1.81 & 2.29 & 2.33 \\
    \hline
     10  & 1.74 & 2.19 & 2.33 & 1.76 & 2.22 & 2.26 \\
    \hline
     13  & 1.68 & 2.10 & 2.24 & 1.69 & 2.13 & 2.18 \\
    \hline
     14  & 1.66 & 2.08 & 2.22 & 1.67 & 2.10 & 2.16 \\
    \hline
  \end{tabular}
 \caption{K-factors for Higgs and pseudo-scalar Higgs boson production cross sections up to ${\rm N^3LO_{\rm SV}}$ 
for different energies at LHC. Here, $m_H=m_A=125$GeV.}
 \label{table:ecmvar}
\end{table}
The QCD corrections to pseudo-scalar Higgs production are found to be
similar to those of the SM Higgs production due to universal infrared
structure of the gluon initiated processes.  We give a numerical
comparison between their K-factors at various orders.  We take
$m_H=m_A=125$ GeV and ignore bottom as well as other light quarks and
electro weak effects for both the cases.  Although the full N$^3$LO
QCD corrections are already available for the SM Higgs boson, for
comparison we take into account only the N$^3$LO$_{\rm SV}$.
Table~\ref{table:ecmvar} contains the K-factors, defined in
Eq.~(\ref{eqn:kf}) up to N$^3$LO$_{\rm SV}$ in QCD for both Higgs and
pseudo-scalar Higgs boson as a function of $\sqrt{S}$.  For this mass
region, the QCD corrections are positive and hence the K-factors
increase with the order in the perturbation theory.  Moreover, these
K-factors, following the line of argument given before, are found to
decrease with $\sqrt{S}$ but they are identical in both the cases. The
difference between the Higgs and the pseudo-scalar cross sections in
their respective K-factors is noticed at the second decimal place
only. At three loop level, ${\rm K^{(3)}}$ is found to be around
$2.4 (2.2)$ for $7(14)$ TeV case.

\begin{table}[h!]
\centering
\begin{tabular}{| c | c | c | c | c | c |  c | c | c |}
    \hline
     \multirow{2}{*}{Mass} & 
     \multicolumn{4}{c|}{SM Higgs} & 
     \multicolumn{4}{c|}{Pseudo-scalar} \\ 
     \cline{2-9}
     {} & LO  & NLO & NNLO & N$^3$LO$_{\rm SV}$ &  LO  & NLO & NNLO & N$^3$LO$_{\rm SV}$\\
    \hline \hline
     124 & 20.32 & 34.08 & 42.76 & 45.60 &  47.02 & 79.46 & 100.03 & 102.54  \\
    \hline
     125 & 20.01 & 33.58 & 42.13 & 44.92 &  46.32 & 78.35 & 98.61 & 101.06  \\
    \hline
     126 & 19.70 & 33.10 & 41.51 & 44.26 &  45.63 & 77.26 & 97.22 & 99.62 \\

    \hline
  \end{tabular}
 \caption{Higgs and pseudo-scalar Higgs cross sections up to ${\rm N^3LO_{\rm SV}}$ for LHC13.}
 \label{table:lhc13-mamh}
\end{table}
The tiny difference between them can be attributed to the presence of
an additional operator present in the effective interaction, namely
${O}_J$ which along with the matching coefficient formally enters from
NNLO onwards for the gluon initiated processes.  For quark anti-quark
initiated processes, this contribution vanishes as the quark flavours
are massless.  The gluon initiated processes involving only ${O}_J$
can contribute at N$^4$LO and beyond.  However, the interference
effects of ${O}_G$ and ${O}_J$ will show up in the gluon initiated
processes at NNLO. Thus, the operator ${O}_J$ has non-zero
contributions at the lowest order namely at two loop level.  However,
the presence of such an interference contribution is found to be very
small and is the main difference between the SM Higgs and the
pseudo-scalar contribution.  The QCD corrections through soft and
collinear gluon emissions for this interference contribution will be
of even higher order and hence will contribute at the three loop level
and beyond.  In table.~\ref{table:lhc13-mamh}, we present the Higgs
and pseudo-scalar Higgs boson production cross sections up to
N$^3$LO$_{\rm SV}$ as a function of the scalar mass around $125$ GeV.
The pseudo-scalar cross section is about twice as big as that of the
Higgs boson and the convergence of perturbation series is good and the
K-factors are roughly the same for both the cases.

\begin{figure*}[htb]
\centerline{
\epsfig{file=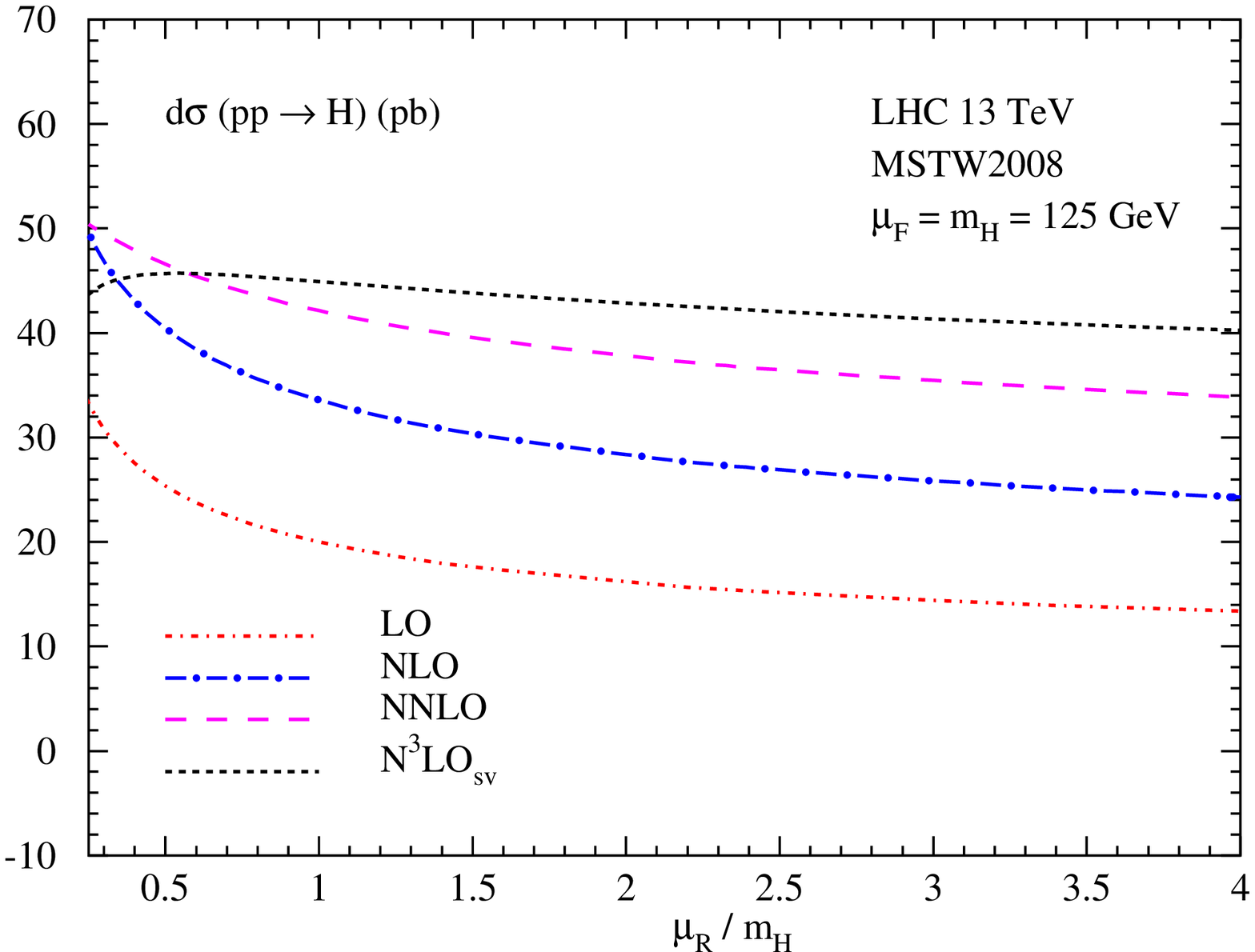,width=8cm,height=6.5cm,angle=0}
\epsfig{file=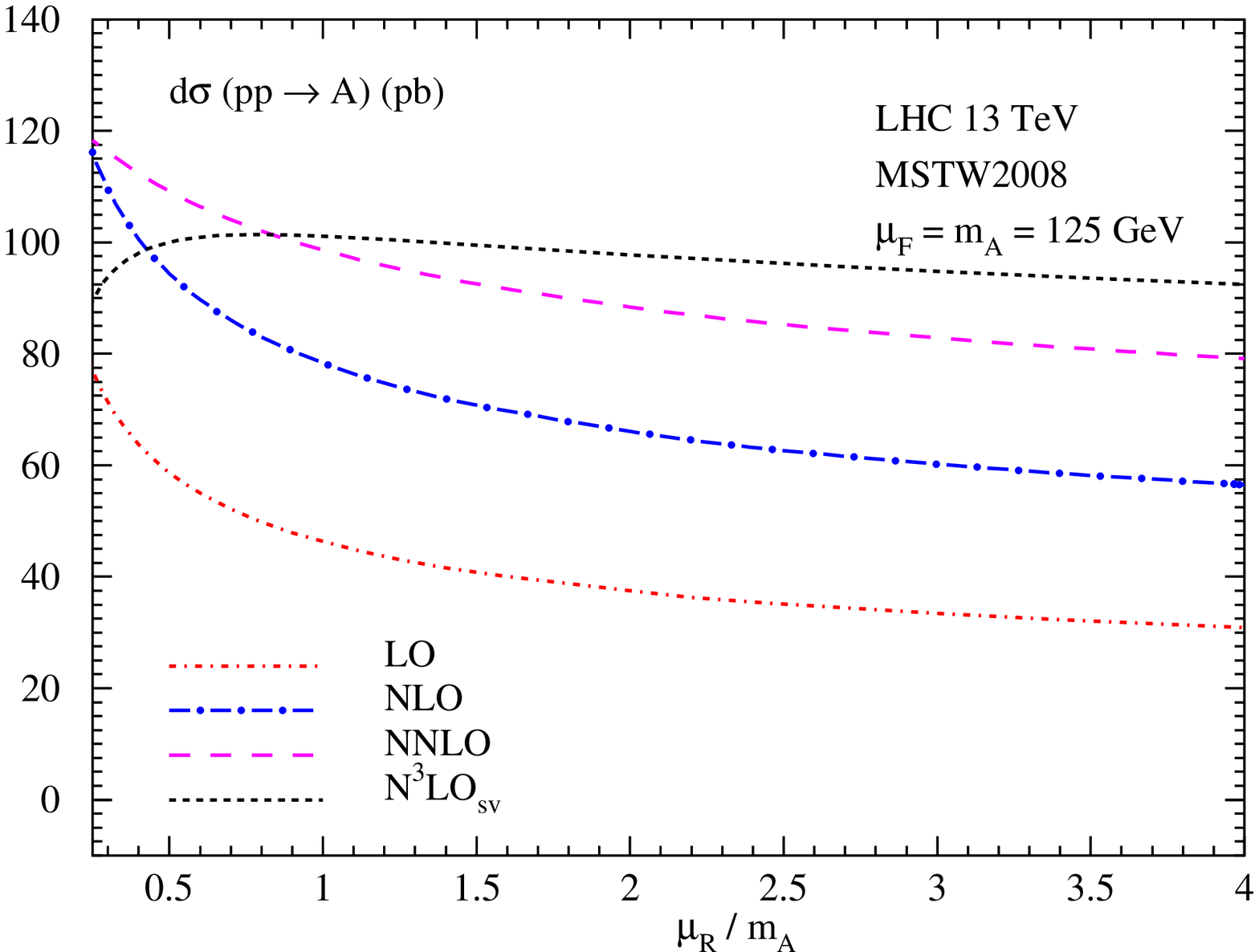,width=8cm,height=6.5cm,angle=0}
}
\caption{\sf Scale uncertainties associated with the pseudo-scalar production cross section 
for LHC13. Variation with $\mu_R$ keeping $\mu_F=m_A$ fixed (left panel). Variation with 
$\mu_F$ keeping $\mu_R = m_A$ fixed (right panel).}
\label{murhps}
\end{figure*}
\begin{figure*}[htb]
\centerline{
\epsfig{file=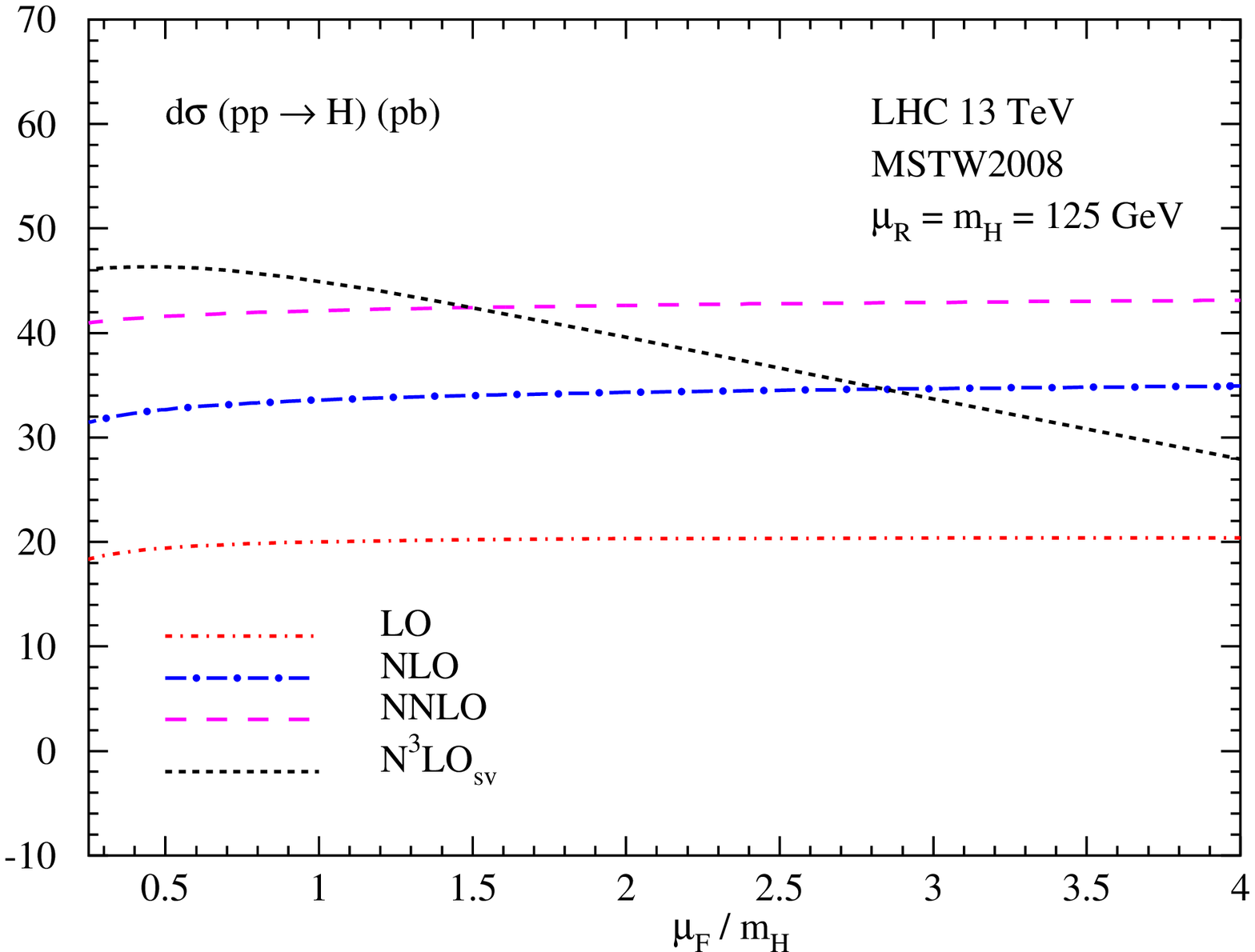,width=8cm,height=6.5cm,angle=0}
\epsfig{file=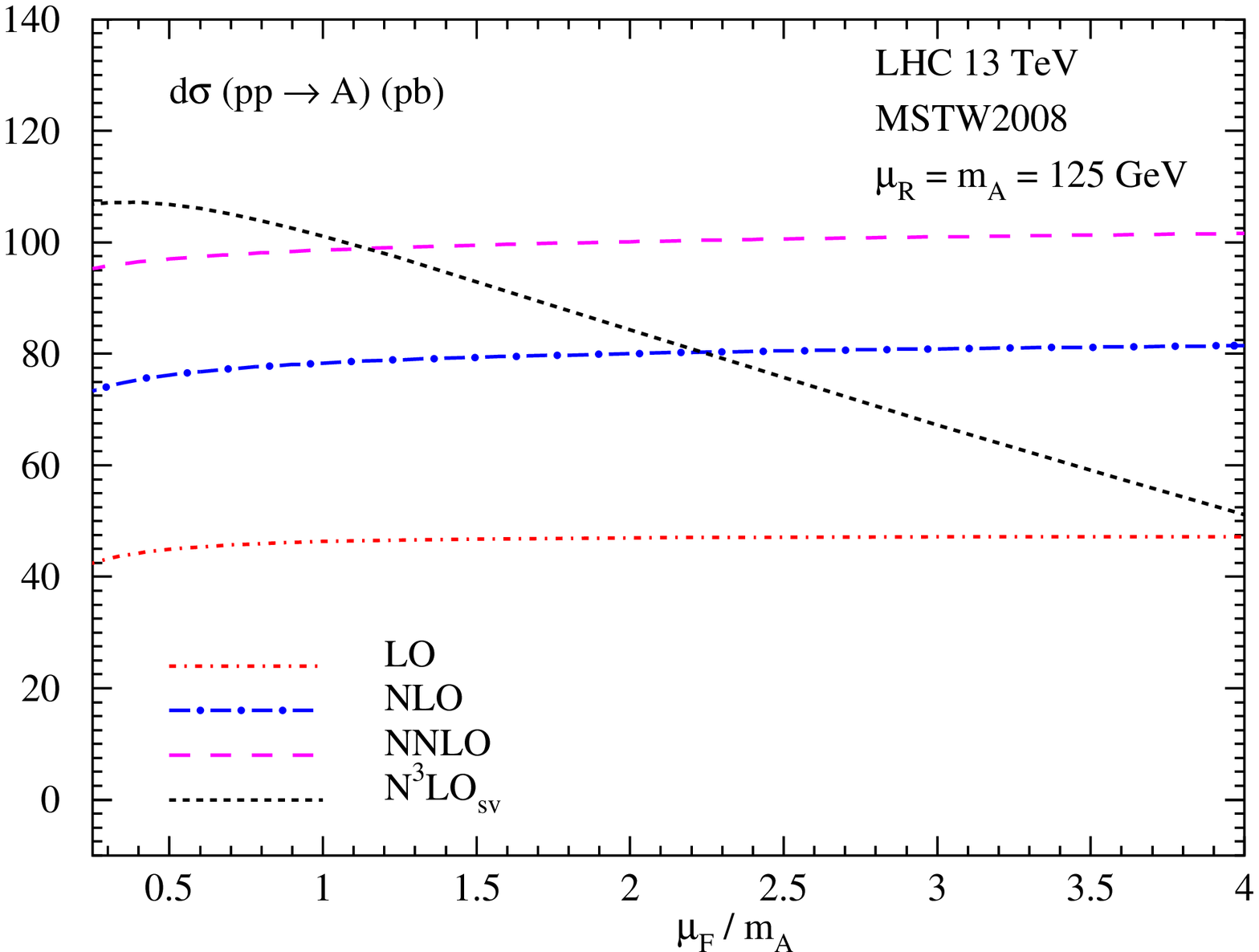,width=8cm,height=6.5cm,angle=0}
}
\caption{\sf Scale uncertainties associated with the pseudo-scalar production cross section 
for LHC13. Variation with $\mu_R$ keeping $\mu_F=m_A$ fixed (left panel). Variation with 
$\mu_F$ keeping $\mu_R = m_A$ fixed (right panel).}
\label{mufhps}
\end{figure*}
\begin{figure*}[htb]
\centerline{
\epsfig{file=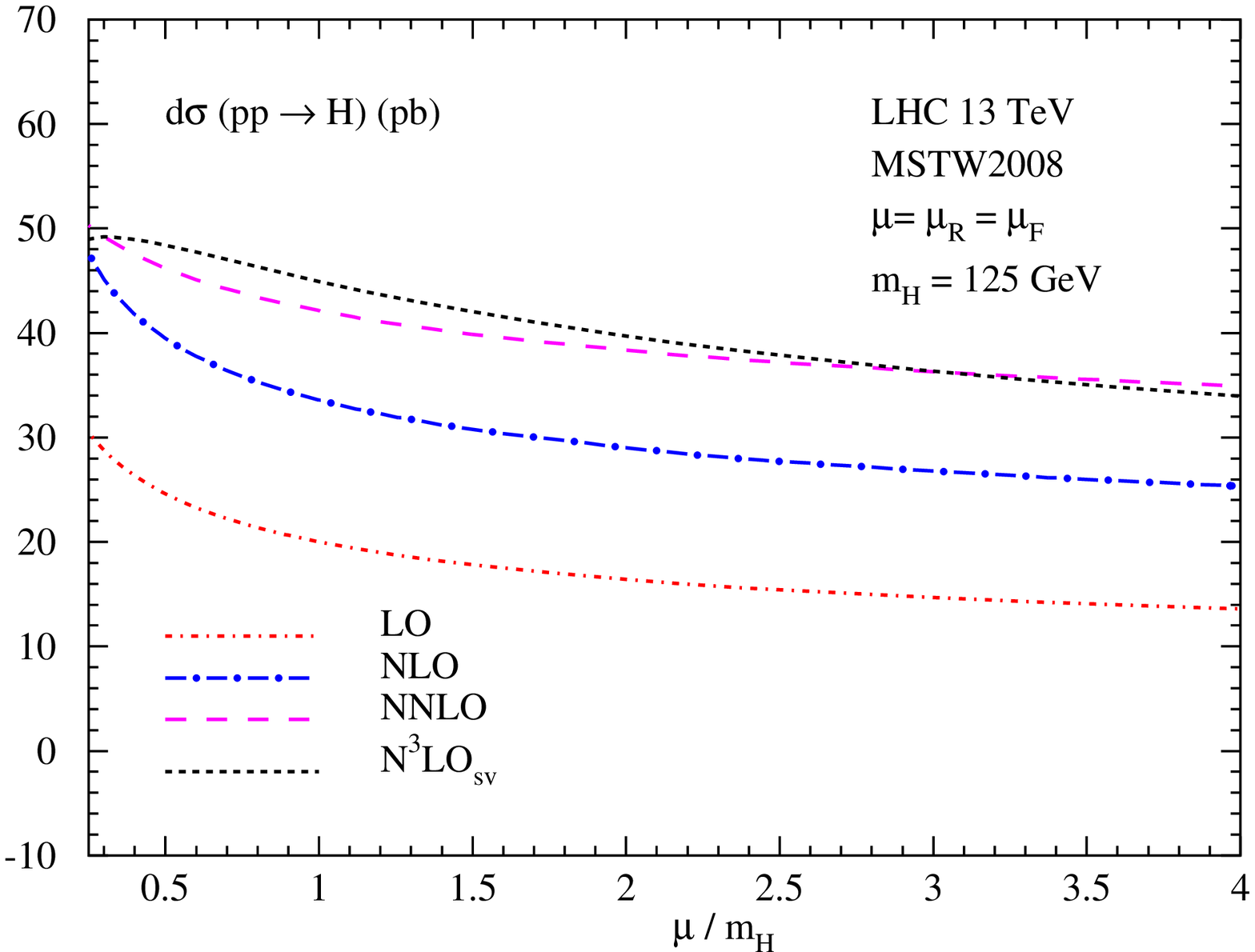,width=8cm,height=6.5cm,angle=0}
\epsfig{file=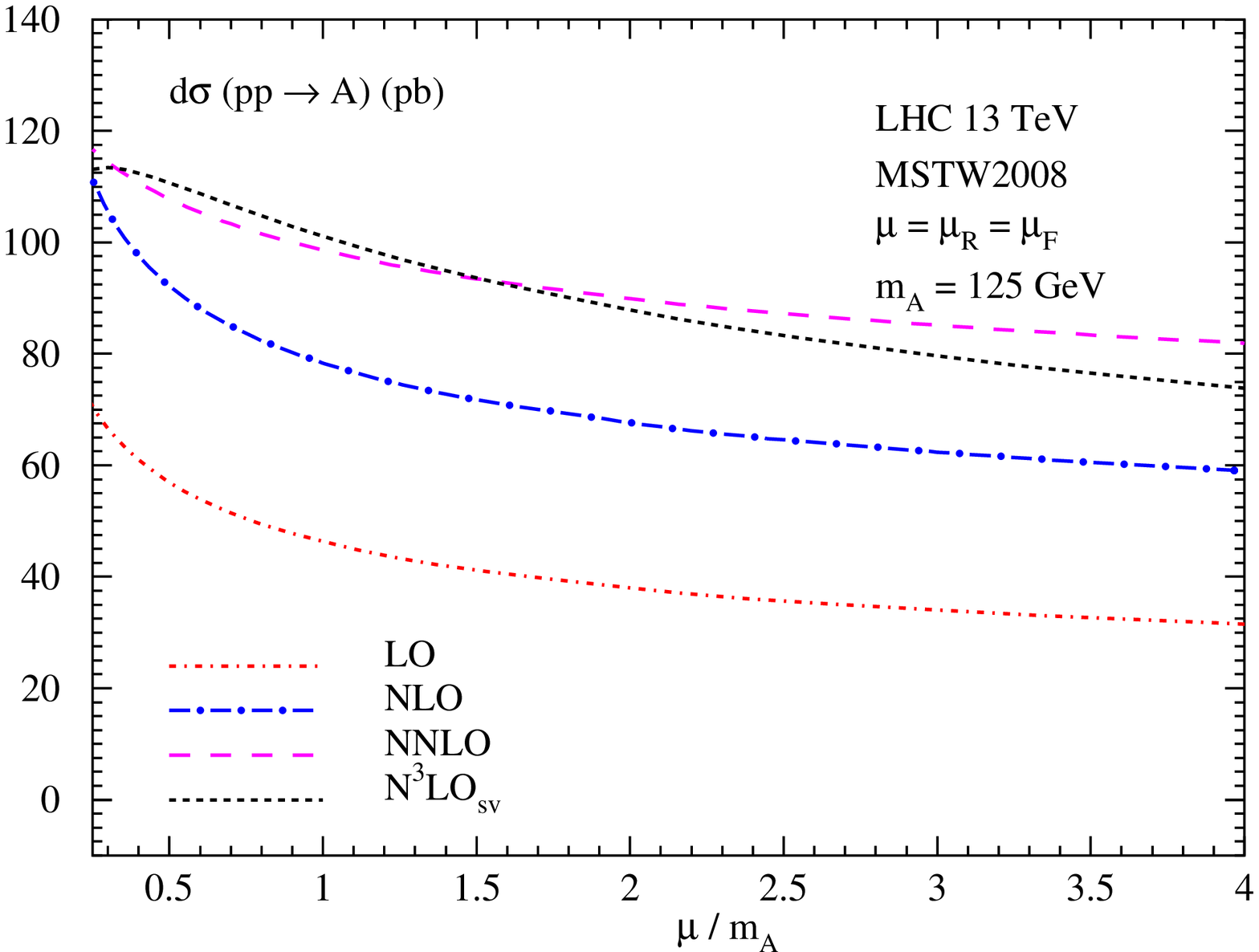,width=8cm,height=6.5cm,angle=0}
}
\caption{\sf Scale uncertainties associated with the pseudo-scalar production cross section 
for LHC13. Variation with $\mu_R$ keeping $\mu_F=m_A$ fixed (left panel). Variation with 
$\mu_F$ keeping $\mu_R = m_A$ fixed (right panel).}
\label{murfhps}
\end{figure*}

Next, we study the renormalisation and factorisation scale variations
of both the cross sections for the production of SM Higgs boson and
pseudo-scalar Higgs boson for $m_H=m_A=125$ GeV by varying them
between $m_A/4$ and $4m_A$. In fig.~\ref{murhps}, the renormalisation
scale uncertainties are given for Higgs boson (left panel) and for
pseudo-scalar boson (right panel), for $\mu_F=m_A=m_H$.  In
fig.~\ref{mufhps}, we present similar results but for the factorisation
scale uncertainties keeping $\mu_R=m_H=m_A$. Moreover, in
fig.~\ref{murfhps}, we present the combined effect by varying
$\mu=\mu_R=\mu_F$. The pattern of the results for the $\mu_R$, $\mu_F$
and the combined variations are similar to the earlier analysis for
$m_A=200$ GeV where the renormalisation scale uncertainties get
stabilised further after including the third order threshold
corrections while the uncertainties due to $\mu_F$ variation get
improved up to NNLO and does not show any improvement at the threshold
N$^3$LO.

\begin{table}[h!]
\centering
\begin{tabular}{| c | c | c | c | c | c | c | }
    \hline
     \multirow{2}{*}{PDF set} & 
     \multicolumn{3}{c|}{SM Higgs} & 
     \multicolumn{3}{c|}{Pseudo-scalar} \\ 
     \cline{2-7}
     {} &  NLO & NNLO & N$^3$LO$_{\rm SV}$ & NLO & NNLO & ${\rm N^3LO_{\rm SV}}$ \\
    \hline \hline
     ABM11  & 33.19 & 39.59 & 41.99 & 77.42 & 92.66 & 94.64 \\
    \hline
     CT10  & 31.79 & 41.84 & 44.67 & 74.15 & 97.94 & 100.44 \\
    \hline
     MSTW2008  & 33.59 & 42.13 & 44.92 & 78.35 & 98.61 & 101.06 \\
    \hline
     NNPDF 23  & 33.55 & 43.01 & 45.87 & 78.26 & 100.70 & 103.19 \\
    \hline
  \end{tabular}
 \caption{PDF uncertainties in the Higgs boson and pseudo-scalar Higgs boson cross sections up to N$^3$LO$_{\rm SV}$ for LHC13 
and for $m_H=m_A=125$GeV.}
 \label{table:lhc13-pdfs}
\end{table}
Since the predictions are sensitive to the choice of parton density
functions, we have estimated the uncertainty resulting from them by
choosing the central fit for various well known PDF sets such
ABM11~\cite{Alekhin:2012ig} , CT10~\cite{Lai:2010vv},
MSTW2008~\cite{Martin:2009iq} and NNPDF23~\cite{Ball:2014uwa}.  For
N$^3$LO$_{\rm SV}$ cross sections, however, we use NNLO PDF sets.  The
corresponding strong coupling constant is directly taken from the {\tt
  LHAPDF}.  In table.~\ref{table:lhc13-pdfs}, we present the SM Higgs
boson and pseudo-scalar Higgs boson production cross sections at NLO,
NNLO and N$^3$LO$_{\rm SV}$ for LHC13.  We find that for NLO, CT10
gives lowest cross section while MSTW2008 gives highest, whereas for
NNLO and N$^3$LO$_{\rm SV}$, ABM11 gives lowest and NNPDF23 gives
highest. The percentage uncertainty arising from PDF sets at any order
is defined as
$(\sigma^A_{\rm max} - \sigma^A_{\rm min})/\sigma^A_{\rm min} \times
100$
where, $\sigma^A_{\rm max}$ and $\sigma^A_{\rm min}$ are the highest
and lowest cross sections at any order obtained from the PDFs
considered, respectively. This PDF uncertainties in the case of Higgs
boson cross sections are about 5.7\% at NLO, 8.6\% at NNLO and 9.2\%
at N$^3$LO$_{\rm SV}$.  For pseudo-scalar production the cross
sections are approximately twice the Higgs cross sections, but the
percentage of PDF uncertainties are almost the same.

\begin{figure*}[htb]
\centerline{
\epsfig{file=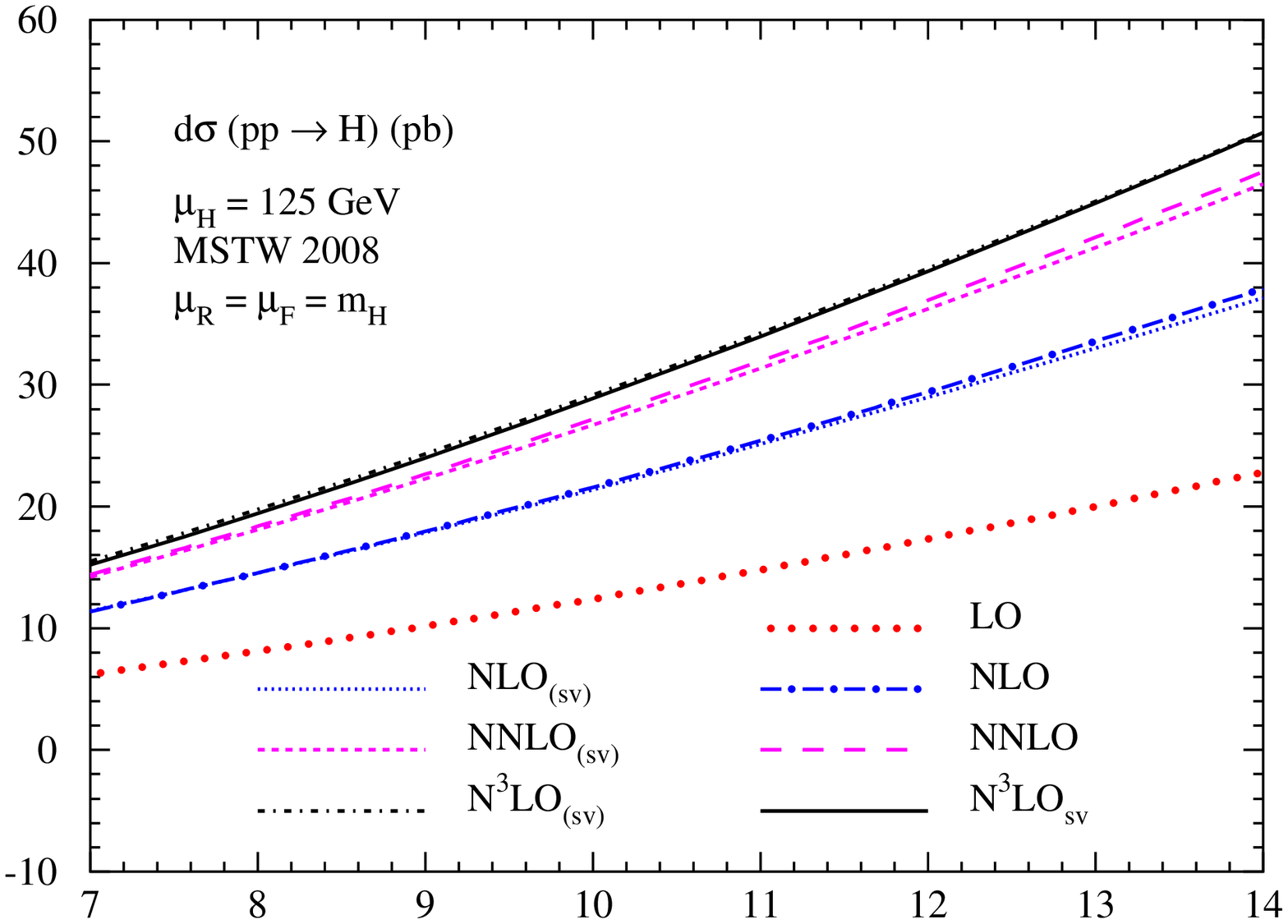,width=8cm,height=6.5cm,angle=0}
\epsfig{file=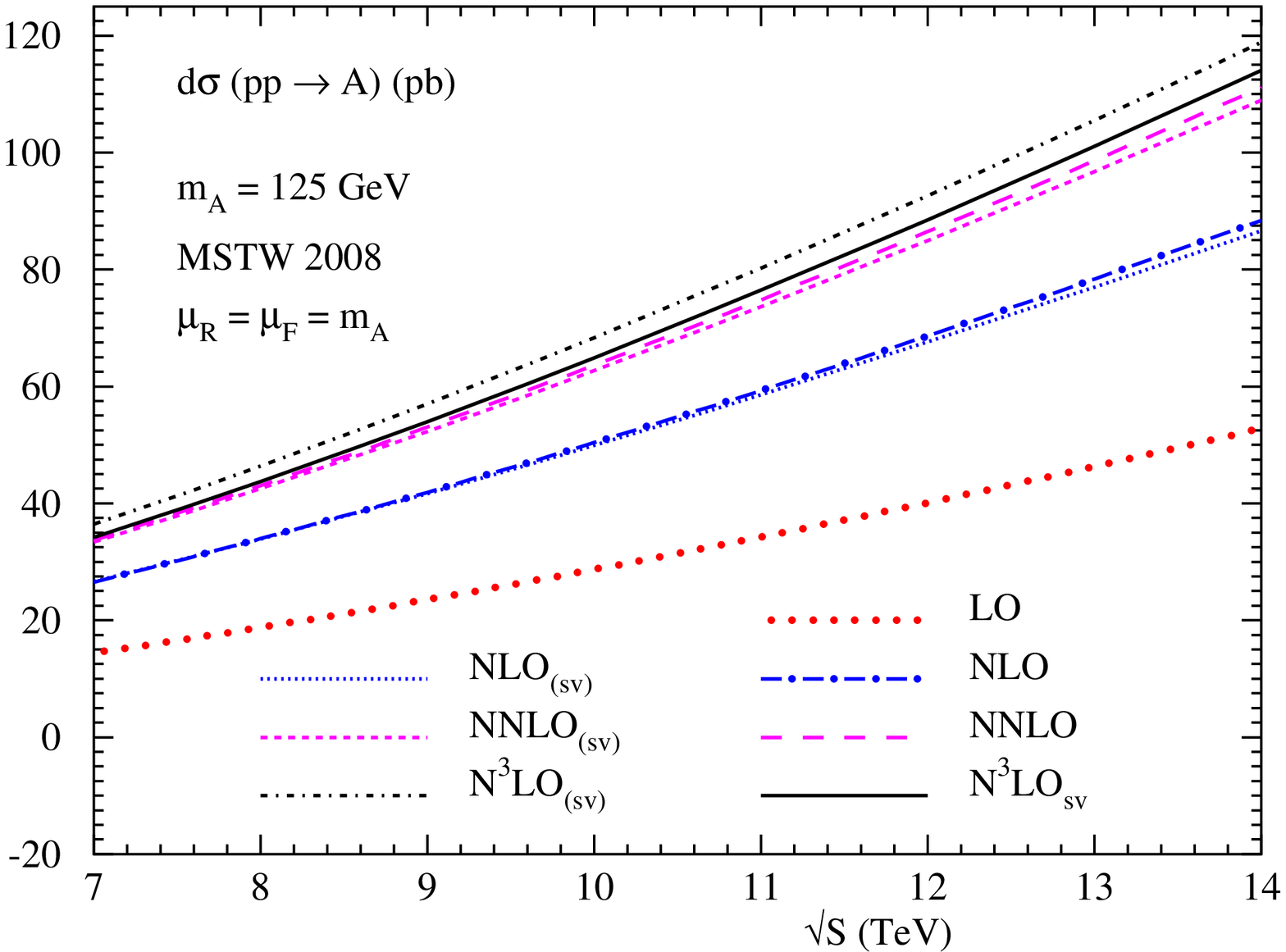,width=8cm,height=6.5cm,angle=0}
}
\caption{\sf Soft-plus-virtual (SVmod) vs fixed order results for Higgs and pseudo-scalar 
Higgs boson production cross section for different energies at LHC.}
\label{svmod}
\end{figure*}

The SV corrections give a rough estimate of the fixed order (FO) QCD
corrections and are often useful in absence of the latter. However,
the relative contribution of these SV corrections to the full FO
results crucially depends on the kinematic region and in some cases on
the process under study.  For the SM Higgs or pseudo-scalar Higgs
boson with a mass of about 125 GeV, it is far from the threshold
region $\tau = m_H^2/S \to 1$ for $\sqrt{S}=13$ TeV. Since, the parton
fluxes corresponding to this mass region are very high, apart from the
threshold logarithms the contributions of the regular terms as well as
of other subprocesses present in the FO corrections are expected to be
reasonably very high.  For Higgs or pseudo-scalar, the prediction at
NLO$_{\rm SV}$ level differs from the LO by only a few percent whereas
the regular terms at NLO contribute significantly and increase LO
prediction by about 70\%. Similar is the case even at NNLO.  Thus the
SV corrections poorly estimate the FO ones, however, if we redefine
the hadron level cross sections without affecting the total cross
sections in such a way that the parton fluxes peak near the threshold
region~\cite{Anastasiou:2014vaa, Ahmed:2014uya, Herzog:2014wja}, then
the SV contributions can be shown to dominate over the regular ones.
This is due to arbitrariness involved in splitting the parton level
cross section in terms of threshold enhanced and regular ones.  Using
a regular function $G(z)$, we can write the hadronic cross section as
\begin{eqnarray}
  \sigma^A(\tau) = \sigma^{A,(0)} \sum_{a,b=q,\bar{q},g} 
  \int_{\tau}^{1} dy ~ G\left({\tau\over y}\right)\Phi_{ab}(y) \left
  ({\Delta^{A}_{ab} \left(\frac{\tau}{y}\right)
  \over G\left({\tau \over y}\right)}\right) 
\label{eqn.tot1}
\end{eqnarray}
where $\Delta(z)/G(z)$ can be decomposed as    
\begin{eqnarray}
\Delta(z)/G(z) = \Delta^{\rm SV}(z) + \tilde \Delta^{\rm hard}(z)
\end{eqnarray}
In the above equation the $\Delta^{\rm SV}$ is independent of $G(z)$
(if $\lim_{z \to 1} G(z) \to 1$) and contains only distributions,
whereas the hard part $\tilde \Delta^{\rm hard}$ is modified due to
$G(z)$.  Hence the SV part of the cross section at the hadron level
depends on the choice of $G(z)$.  For the peculiar choice
$G(z) = z^2$, the $\Delta^{\rm SV}$ dominates over
$\tilde \Delta^{\rm hard}$ in such a way that almost the entire NLO
and NNLO corrections (Eq.~(\ref{eqn.tot1})) results from
$\Delta^{\rm SV}$ alone.  As was noted earlier $G(z)=1$ corresponds to
the standard SV contribution.  Note that the flux $\Phi_{ab}$ is
modified to $\Phi^{\rm mod}_{ab}(y)=\Phi_{ab}(y) G(\tau/y)$ which is
responsible for this behaviour.  We may denote the SV cross sections
thus obtained with these modified fluxes as
${\rm NLO_{(sv)}}$, ${\rm NNLO_{(sv)}}$ and ${\rm N^3LO_{(sv)}}$
while those obtained with the normal fluxes as ${\rm NLO_{sv}}$,
${\rm NNLO_{sv}}$ and ${\rm N^3LO_{sv}}$.  In fig.~\ref{svmod}, we
depict the comparison between the SV cross sections obtained from the
modified parton fluxes using $G(z)=z^2$ and the normal fixed order
results that are obtained from the standard parton fluxes, for both
the SM Higgs boson (left panel) and the pseudo-scalar (right panel).
We notice that the SV results are significantly closer to the
corresponding fixed order ones. Incidentally, this agreement is good
for NLO as well as for NNLO where different subprocesses appear, and
also for several values of $\sqrt{S}$ where the integration range over
the parton fluxes is different. While this could be purely accidental,
this good agreement might hint some subtle aspect hidden and might be
useful in the phenomenology.

Motivated by the above observation, one can convolute the perturbative
coefficients $\Delta_{\rm SV}^{(3)}$ with the modified parton fluxes
$\Phi^{\rm mod}_{ab}(y)$ for the choice of $G(z)=z^2$ to get
${\rm N^3LO_{(sv)}}$ which could approximate the full $\rm N^3LO$
result. This way, we present in fig.~\ref{svmod}, the SV corrections
obtained using $G(z)=1$ and $G(z)=z^2$ for Higgs as well as
pseudo-scalar Higgs boson productions.

\section{Conclusions}
\label{sec:conc}

In this paper, using the recently available pseudo-scalar form factors
up to three loops and the third order soft function from the real
radiations, a complete N$^3$LO threshold correction to the production
of pseudo-scalar at the LHC has been obtained.  The
computation is performed using $z$ space representation of resummed
cross section.  We have exploited the universal structure of soft
function that appears in scalar Higgs boson production at the LHC.  We found
that the singularities resulting from soft and collinear regions in
the virtual diagrams cancel against those from the universal soft
functions as well as from mass factorisation kernels.  Using our
approach, we have also computed the process dependent coefficient that
appears in the threshold resummed cross section.  This will be useful
for resummed predictions at N$^3$LL in QCD.  Using threshold corrected
N$^3$LO results, we have presented a detailed phenomenological study
of the pseudo-scalar production at the LHC for various
center of mass energies as a function of its mass.  While the third
order corrections are small, they play an important role in reducing
the theoretical uncertainty resulting from renormalisation scale.  In
addition, we have made a detailed comparison against scalar Higgs
boson production and found their corrections are very close to each
other confirming the universal behaviour of the QCD effects even
though the operators responsible for their interactions with gluons
are very different.

\section*{Acknowledgement}

We sincerely thank Thomas Gehrmann for fruitful discussions. We would
also like to thank Roman N. Lee for useful discussions and timely help.

\bibliography{main} \bibliographystyle{utphysM}
  
\end{document}